\documentclass[twocolumn]{aastex631}

\newcommand{\sfrunit}{M$_{\odot}$~yr$^{-1}$}
\usepackage{hyperref}  % so TAH can link github & gdrive folder using \href{}
\definecolor{notes}{HTML}{5C9384}
\newcommand{\jdrp}{\texttt{jwst}}

\usepackage{multirow}

\begin{document}
% SGAS and Megasaura stuff:
\newcommand{\megasaura}{M\textsc{eg}a\textsc{S}a\textsc{ura}}
\newcommand{\megasauralong}{The Magellan Evolution of Galaxies Spectroscopic and Ultraviolet Reference Atlas}
\newcommand{\rcsohthree}{RCS-GA~032727$-$132609}
\newcommand{\stripleohfour}{SGAS~J000451.7$-$010321}
\newcommand{\sohohthreethree}{SGAS~J003341.5$+$024217}
\newcommand{\sohoneoheight}{SGAS~J010842.2$+$062444}
\newcommand{\sohninehundred}{SGAS~J090003.3$+$223408} 
\newcommand{\sohninefiftyseven}{SGAS~J095738.7$+$050929}
\newcommand{\stenfifty}{SGAS~J105039.6$+$001730}
\newcommand{\stwelvetwentysix}{SGAS J122651.3$+$215220}
\newcommand{\sfourteentwentynine}{SGAS~J142954.9$+$120239}
\newcommand{\sfourteenfiftyeight}{SGAS~J145836.1$-$002358} 
\newcommand{\sfifteentwentyseven}{SGAS~J152745.1$+$065219}
\newcommand{\stwentyoneeleven}{SGAS~J211118.9$-$011431}
\newcommand{\stwentytwofortythree}{SGAS~J224324.2$-$093508}  

% Types of stacked megasaura spectra:
\newcommand{\shapenorm}{shape-normalized}
\newcommand{\pivot}{$\lambda_{pivot}$-normalized}
\newcommand{\deredpivot}{dereddened $\lambda_{pivot}$-normalized}
\newcommand{\snine}{S99-normalized}
%
% Missions
\newcommand{\jwst}{\textit{JWST}}
\newcommand{\hst}{\textit{HST}}
%
%latex crap 
\newcommand{\etal}{et~al.}
\newcommand\revisedApJ[1]{\textcolor{blue}{\textbf{#1}}}   % good for revising text in ApJ resubmissions
%\newcommand{\la}{\lesssim}  %only needed when not in aastex
%\newcommand{\ga}{\gtrsim}   %only needed when not in aastex 
%\newcommand{\degr}{\arcdeg}
%
% Astronomy definitions
\newcommand{\Lsun}{L$_{\odot}$}
\newcommand{\Msun}{M$_{\odot}$}
\newcommand{\Lsol}{L$_{\odot}$}
\newcommand{\Msol}{M$_{\odot}$}
\newcommand{\Zsol}{Z$_{\odot}$}
\newcommand{\Mstellar}{M$_{*}$}
\newcommand{\Mbh}{M$_{BH}$}
\newcommand{\Lbh}{L$_{BH}$}
%
% ASTRONOMY JARGON
\newcommand{\Teff}{\hbox{T$_{eff}$}}
\newcommand{\Mup}{\hbox{M$_{up}$}}
\newcommand{\ba}{$b/a$}
\newcommand{\ebv}{$E(B - V)$}
\newcommand{\Hunits}{km~s$^{-1}$~Mpc$^{-1}$}
%
% cgs and other unit abbreviations
\newcommand{\cgsflux}{erg~s$^{-1}$~cm$^{-2}$}
\newcommand{\cgsflam}{erg~s$^{-1}$~cm$^{-2}$~\AA$^{-1}$}
\newcommand{\cgsfnu}{erg~s$^{-1}$~cm$^{-2}$~Hz$^{-1}$}
\newcommand{\cmsq}{\hbox{cm$^{-2}$}}
\newcommand{\ergss}{erg~s$^{-1}$}
\newcommand{\kms}{\hbox{km~s$^{-1}$}}
\newcommand{\cc}{\hbox{cm$^{-3}$}}
\newcommand{\peryr}{\hbox{yr$^{-1}$}}
%
% Bands and filters
\newcommand{\hard}{$2$--$8$~keV}
\newcommand{\soft}{$0.5$--$2$~keV}
%
% Atomic lines and line ratios
\newcommand{\ciiidoublet}{[C~III] 1907, C~III] 1909~\AA}
\newcommand{\ciiidoubletnoAA}{[C~III] 1907, C~III] 1909}
\newcommand{\oiidoublet}{[O II]~3727, 3729~\AA}
\newcommand{\oiiidoublet}{[O III]~4959, 5007~\AA}
\newcommand{\oiiiuv}{[O III]~1660, 1666~\AA}
\newcommand{\siiiiuv}{Si III~1883~\AA}
\newcommand{\hab}{H$\alpha$/H$\beta$}
\newcommand{\mgii}{Mg~II~2796, 2803~\AA}
\newcommand{\civ}{C~IV 1548, 1551~\AA}
\newcommand{\lya}{Ly~$\alpha$}
\newcommand{\ArII}{\hbox{{\rm Ar}\kern 0.1em{\sc ii}}}
\newcommand{\ArIII}{\hbox{{\rm Ar}\kern 0.1em{\sc iii}}}
\newcommand{\CIV}{\hbox{{\rm C}\kern 0.1em{\sc iv}}}
\newcommand{\HI}{\hbox{{\rm H}\kern 0.1em{\sc i}}}
\newcommand{\HII}{\hbox{{\rm H}\kern 0.1em{\sc ii}}}
\newcommand{\HeI}{\hbox{{\rm He}\kern 0.1em{\sc i}}}
\newcommand{\HeII}{\hbox{{\rm He}\kern 0.1em{\sc ii}}}
\newcommand{\NII}{\hbox{{\rm N}\kern 0.1em{\sc ii}}}
\newcommand{\OI}{\hbox{{\rm O}\kern 0.1em{\sc i}}}
\newcommand{\OII}{\hbox{{\rm O}\kern 0.1em{\sc ii}}}
\newcommand{\OIII}{\hbox{{\rm O}\kern 0.1em{\sc iii}}}
\newcommand{\OIIlong}{{\rm O}\kern 0.1em{\sc ii}~$\lambda 3727$} 
\newcommand{\FeII}{\hbox{{\rm Fe}\kern 0.1em{\sc ii}}}
\newcommand{\NeII}{\hbox{{\rm Ne}\kern 0.1em{\sc ii}}}
\newcommand{\NeIII}{\hbox{{\rm Ne}\kern 0.1em{\sc iii}}}
\newcommand{\NeV}{\hbox{{\rm Ne}\kern 0.1em{\sc v}}}
\newcommand{\SII}{\hbox{{\rm S}\kern 0.1em{\sc ii}}}
\newcommand{\SIII}{\hbox{{\rm S}\kern 0.1em{\sc iii}}}
\newcommand{\SIV}{\hbox{{\rm S}\kern 0.1em{\sc iv}}}
\newcommand{\SiIV}{\hbox{{\rm Si}\kern 0.1em{\sc iv}}}
\newcommand{\MgII}{\hbox{{\rm Mg}\kern 0.1em{\sc ii}}}
\newcommand{\Halpha}{\hbox{{\rm H}\kern 0.1em$\alpha$}}
\newcommand{\Hbeta}{\hbox{{\rm H}\kern 0.1em$\beta$}}
\newcommand{\Heopta}{\hbox{{\rm He}\kern 0.1em{\sc i}}~$6678$}
\newcommand{\Heoptb}{\hbox{{\rm He}\kern 0.1em{\sc i}}~$5876$}
\newcommand{\Heoptc}{\hbox{{\rm He}\kern 0.1em{\sc i}}~$4471$}
\newcommand{\Brgam}{\hbox{{\rm Br}\kern 0.1em$\gamma$}}
\newcommand{\Brten}{\hbox{{\rm Br}\kern 0.1em$10$}}
\newcommand{\Breleven}{\hbox{{\rm Br}\kern 0.1em$11$}}
\newcommand{\HeIh}{\hbox{{\rm He}\kern 0.1em{\sc i}}~$1.7$~{\micron}}
\newcommand{\HeIk}{\hbox{{\rm He}\kern 0.1em{\sc i}}~$2.06$~{\micron}}
%
% flux and luminosity Ratios
\newcommand{\hs}{$2$-$8$~keV$/0.5$-$2$~keV}
\newcommand{\xmips}{$24$~\micron$/2$-$8$~keV}
\newcommand{\fxr}{f$_X/$f$_R$}
\newcommand{\logfxr}{$\log$ (f$_X/$f$_R$)}
\newcommand{\longfxr}{f$_{\nu=R}/$f$_{\nu=2-8~keV}$}
\newcommand{\nufnu}{$\nu f_{\nu}$}
\newcommand{\fnu}{$f_{\nu}$}

%
% Instruments and Spacecraft
\newcommand{\spitzer}{\emph{Spitzer}}
\newcommand{\chandra}{\emph{Chandra}}

% compressed lists (good for proposals)
\newcommand{\squishlist}{
   \begin{list}{$\bullet$}
    { \setlength{\itemsep}{0pt}      \setlength{\parsep}{1pt}
      \setlength{\topsep}{3pt}       \setlength{\partopsep}{0pt}
      \setlength{\leftmargin}{1.5em} \setlength{\labelwidth}{1em}
      \setlength{\labelsep}{0.5em} } }
\newcommand{\squishend}{
    \end{list}  }  % add Jane Rigby's shortcuts

\title{JWST Early Release Science Program TEMPLATES: Targeting Extremely Magnified Panchromatic Lensed Arcs and their Extended Star formation}

% TAH testing for JRR, moving affilliations to end of document
% to clean up first page (so abstract is actually on first page)
\suppressAffiliations

\correspondingauthor{Jane R.~Rigby}
\email{Jane.Rigby@nasa.gov}

% PI and Co-PI first

\author[0000-0002-7627-6551]{Jane R. Rigby}
\affiliation{Astrophysics Science Division, Code 660, NASA Goddard Space Flight Center, 8800 Greenbelt Rd., Greenbelt, MD 20771, USA}

\author[0000-0001-7192-3871]{Joaquin D. Vieira}	
\affiliation{Department of Astronomy, University of Illinois Urbana-Champaign, 1002 West Green St., Urbana, IL 61801, USA}	
\affiliation{Department of Physics, University of Illinois Urbana-Champaign, 1110 West Green St., Urbana, IL 61801, USA}	
\affiliation{Center for AstroPhysical Surveys, National Center for Supercomputing Applications, 1205 West Clark Street, Urbana, IL 61801, USA}

% Substantial contributors here:

\author[0000-0001-7946-557X]{Kedar A. Phadke}	
\affiliation{Department of Astronomy, University of Illinois Urbana-Champaign, 1002 West Green St., Urbana, IL 61801, USA}	
\affiliation{Center for AstroPhysical Surveys, National Center for Supercomputing Applications, 1205 West Clark Street, Urbana, IL 61801, USA}

\author[0000-0001-6251-4988]{Taylor A.\ Hutchison}	
\altaffiliation{NASA Postdoctoral Fellow}
\affiliation{Astrophysics Science Division, Code 660, NASA Goddard Space Flight Center, 8800 Greenbelt Rd., Greenbelt, MD 20771, USA}

\author[0000-0003-1815-0114]{Brian Welch}	
\affiliation{Astrophysics Science Division, Code 660, NASA Goddard Space Flight Center, 8800 Greenbelt Rd., Greenbelt, MD 20771, USA}
\affiliation{Department of Astronomy, University of Maryland, College Park, MD 20742, USA}	
\affiliation{Center for Research and Exploration in Space Science and Technology, NASA/GSFC, Greenbelt, MD 20771}

\author[0000-0002-4657-7679]{Jared Cathey}	
\affiliation{Department of Astronomy, University of Florida, 211 Bryant Space Sciences Center, Gainesville, FL 32611 USA}
\author[0000-0003-3256-5615]{Justin~S.~Spilker}	
\affiliation{Department of Physics and Astronomy and George P. and Cynthia Woods Mitchell Institute for Fundamental Physics and Astronomy, Texas A\&M University, 4242 TAMU, College Station, TX 77843-4242, US}

\author[0000-0002-0933-8601]{Anthony H.~Gonzalez}	
\affil{Department of Astronomy, University of Florida, 211 Bryant Space Sciences Center, Gainesville, FL 32611 USA}

% author list is alphabetical after here
\author[0009-0003-3123-4897]{Prasanna Adhikari}
\affiliation{Department of Physics, University of Cincinnati, Cincinnati, OH 45221, USA}

\author[0000-0002-6290-3198]{M.~Aravena}
\affil{N\'{u}cleo de Astronom{\'i}a de la Facultad de Ingenier\'{i}a y Ciencias, Universidad Diego Portales, Av. Ej\'{e}rcito Libertador 441, Santiago, Chile}

\author[0000-0003-1074-4807]{Matthew B. Bayliss}
\affiliation{Department of Physics, University of Cincinnati, Cincinnati, OH 45221, USA}

\author[0000-0002-3272-7568]{Jack E. Birkin}
\affiliation{Department of Physics and Astronomy and George P. and Cynthia Woods Mitchell Institute for Fundamental Physics and Astronomy, Texas A\&M University, 4242 TAMU, College Station, TX 77843-4242, US}

\author[0009-0008-8439-7442]{Emmy Bursk}
\affiliation{Department of Physics, University of Cincinnati, Cincinnati, OH 45221, USA}

\author[0000-0002-8487-3153]{Scott C.\ Chapman}
\affiliation{Department of Physics and Atmospheric Science, Dalhousie University, Halifax, NS, B3H 4R2, Canada}
\affiliation{NRC Herzberg Astronomy and Astrophysics, 5071 West Saanich Rd, Victoria, BC, V9E 2E7, Canada}
\affiliation{Department of Physics and Astronomy,  University of British Columbia, Vancouver, BC, V6T1Z1, Canada}

\author[0000-0003-2200-5606]{H\r{a}kon Dahle} \affiliation{Institute of Theoretical Astrophysics, University of Oslo, P.O. Box 1029, Blindern, NO-0315 Oslo, Norway}

\author[0009-0007-6157-7398]{Lauren A. Elicker}
\affiliation{Department of Physics, University of Cincinnati, Cincinnati, OH 45221, USA}

\author[0000-0002-3365-8875]{Travis C. Fischer}
\affiliation{AURA for ESA, Space Telescope Science Institute, 3700 San Martin Drive, Baltimore, MD 21218, USA} 

\author[0000-0001-5097-6755]{Michael K. Florian}
\affiliation{Steward Observatory, University of Arizona, 933 North Cherry Ave., Tucson, AZ 85721, USA}

\author[0000-0003-1370-5010]{Michael D.~Gladders}
\affiliation{Kavli Institute for Cosmological Physics, University of Chicago, 5640 South Ellis Avenue, Chicago, IL 60637, USA}

\author[0000-0003-4073-3236]{Christopher C. Hayward}
\affiliation{Center for Computational Astrophysics, Flatiron Institute, 162 Fifth Avenue, New York, NY 10010, USA}

\author[0000-0002-6823-655X]{Rose Hewald}
\affiliation{Department of Physics \& Astronomy, University of Notre Dame, Notre Dame, IN 46556, USA}

\author[0009-0008-5151-7639]{Lily A.\,Kettler}
\affiliation{Department of Astronomy, University of Illinois Urbana-Champaign, 1002 West Green St., Urbana, IL 61801, USA}	

\author[0000-0002-3475-7648]{Gourav Khullar}
\affiliation{Department of Physics and Astronomy, and PITT PACC, University of Pittsburgh, Pittsburgh, PA 15260, USA}

\author[0000-0002-6787-3020]{Seonwoo Kim}	
\affiliation{Department of Astronomy, University of Illinois  Urbana-Champaign, 1002 West Green St., Urbana, IL 61801, USA}	

\author[0000-0002-9402-186X]{David R.\ Law}
\affiliation{Space Telescope Science Institute, 3700 San Martin Drive, Baltimore, MD, 21218, USA}

\author[0000-0003-3266-2001]{Guillaume Mahler}
\affiliation{STAR Institute, Quartier Agora - All\'ee du six Ao\^ut, 19c B-4000 Li\`ege, Belgium}

\author[0000-0002-9226-5350]{Sangeeta Malhotra}
\affiliation{Astrophysics Science Division, Code 665, NASA Goddard Space Flight Center, 8800 Greenbelt Rd., Greenbelt, MD 20771, USA}

\author[0000-0001-7089-7325]{Eric J.\,Murphy}
\affiliation{National Radio Astronomy Observatory, 520 Edgemont Road, Charlottesville, VA 22903, USA}

\author[0000-0002-7064-4309]{Desika Narayanan}
\affil{Department of Astronomy, University of Florida, 211 Bryant Space Sciences Center, Gainesville, FL 32611 USA}
\affil{University of Florida Informatics Institute, 432 Newell Drive, CISE Bldg E251, Gainesville, FL 32611}
\affil{Cosmic Dawn Center at the Niels Bohr Institute, University of Copenhagen and DTU-Space, Technical University of Denmark}

\author[0000-0002-4606-4240]{Grace M. Olivier}
\affiliation{Department of Physics and Astronomy and George P. and Cynthia Woods Mitchell Institute for Fundamental Physics and Astronomy, Texas A\&M University, 4242 TAMU, College Station, TX 77843-4242, US}

\author[0000-0002-1501-454X]{James E. Rhoads}
\affiliation{Astrophysics Science Division, Code 667, NASA Goddard Space Flight Center, 8800 Greenbelt Rd., Greenbelt, MD 20771, USA}

\author[0000-0002-7559-0864]{Keren Sharon}
\affiliation{Department of Astronomy, University of Michigan, 1085 S. University Ave, Ann Arbor, MI 48109, USA}

\author[0000-0001-6629-0379]{Manuel Solimano}
\affiliation{Instituto de Estudios Astrof\'{\i}sicos, Facultad de Ingenier\'{\i}a y Ciencias, Universidad Diego Portales, Avenida Ej\'ercito Libertador 441, Santiago, Chile. [C\'odigo Postal 8370191]}

\author[0009-0005-1043-0615]{Athish Thiruvengadam}
\affiliation{Department of Physics, University of Illinois  Urbana-Champaign, 1110 West Green St., Urbana, IL 61801, USA}
\affiliation{Department of Astronomy, University of Illinois Urbana-Champaign, 1002 West Green St., Urbana, IL 61801, USA}	

\author[0000-0002-0786-7307]{David Vizgan}
\affiliation{Department of Astronomy, University of Illinois Urbana-Champaign, 1002 West Green St., Urbana, IL 61801, USA}	

\author[0009-0009-9873-833X]{Nikolas Younker}
\affiliation{Department of Physics, University of Cincinnati, Cincinnati, OH 45221, USA}

\collaboration{1000}{(TEMPLATES collaboration)}
% Stop truncating the author list, Latex!  -JR

%% AASTeX 6.31 has the new \collaboration and \nocollaboration commands to
%% provide the collaboration status of a group of authors. These commands 
%% can be used either before or after the list of corresponding authors. The
%% argument for \collaboration is the collaboration identifier. Authors are
%% encouraged to surround collaboration identifiers with ()s. The 
%% \nocollaboration command takes no argument and exists to indicate that
%% the nearby authors are not part of surrounding collaborations.

%% Mark off the abstract in the ``abstract'' environment. 
\begin{abstract}
This paper gives an overview of TEMPLATES, a JWST Early Release Science program that targeted four extremely bright,  gravitationally lensed galaxies: two extremely dusty, two with low attenuation, as templates for galaxy evolution studies with JWST.  TEMPLATES obtains a common set of spectral diagnostics for these $1.3 \le z \le 4.2$ galaxies, in particular H$\alpha$, Paschen$\alpha$, and the rest-frame optical and near-infrared continua. In addition, two of the four targets have JWST coverage of [O~III]~5007\AA~ and H$\beta$; the other two targets have have JWST coverage of PAH 3.3$\mu$m and complementary ALMA data covering the [C II] 158 micron emission line.  The science goals of TEMPLATES are to demonstrate attenuation-robust diagnostics of star formation, map the distribution of star formation, compare the young and old stellar populations, and measure the physical conditions of star formation and their spatial variation across the galaxies.  In addition, TEMPLATES has technical goals to establish best practices for the Integral Field Units (IFU) within the NIRSpec and MIRI instruments, both in terms of observing strategy and in terms of data reduction.  The paper describes TEMPLATES's observing program, scientific and technical goals,  data reduction methods, and deliverables, including high-level data products and data reduction cookbooks.
\end{abstract}

%% Keywords should appear after the \end{abstract} command. 
%% The AAS Journals now uses Unified Astronomy Thesaurus concepts:
%% https://astrothesaurus.org
%% You will be asked to selected these concepts during the submission process
%% but this old "keyword" functionality is maintained in case authors want
%% to include these concepts in their preprints.

\section{Introduction} \label{sec:intro}
This is the overview paper for TEMPLATES, a 54~hr JWST observing program that was part of the Director's Discretionary Early Release Science (DD-ERS) initiative.  TEMPLATES is a contrived acronym for ``Targeting Extremely Magnified Panchromatic Lensed Arcs and Their Extended Star formation'';  the program ID (PID) is 1355.
TEMPLATES pairs the exquisite spatial resolution and multiplexed spectroscopic capabilities of JWST with the natural telescopes that are strong gravitational lenses.
The  science goals of TEMPLATES are to spatially resolve the star formation in four gravitationally lensed galaxies, and to characterize the physical conditions of star formation across a broad range of dust obscuration.   The website for the program is \href{https://sites.google.com/view/jwst-templates/}{sites.google.com/view/jwst-templates}, and the github repository is \href{https://github.com/JWST-Templates}{github.com/JWST-Templates}.

The Early Release Science (ERS) initiative is a set of 13 JWST observing programs, totalling $\sim 450$~hr of Director's Discretionary Time (DDT), that were selected in 2017 through competitive peer review.  In addition to the usual selection criteria of scientific merit, ERS programs were also solicited to serve the scientific user community: to obtain representative datasets early in the mission lifetime, to support community preparation of Cycle 2 and 3 proposals, to engage a broad cross-section of the astronomical community, and to help users become familiar with JWST data and JWST's scientific capabilities.   To support these community service goals, the ERS programs were preferentially scheduled early in the first year of  JWST science operations.

TEMPLATES accomplishes the ERS goals by conducting compelling extragalactic science, by generating data cubes and derived data products with high signal-to-noise ratio and high dynamic range, and by exercising four science instrument modes and a wide variety of setups within those modes:  four NIRSpec IFS grating/filter setups,  six NIRCam imaging filters, 
 seven MIRI imaging filters, and two of the three MIRI MRS grating settings.
TEMPLATES' key deliverables include science-ready data products and high-level science products, lens models, and Python notebooks that document how we reduced the data, by using a combination of the \jdrp\ pipeline, third party tools, and our own custom steps.  Our goal in releasing these notebooks is to enable the user community to efficiently process their own datasets, particularly in the spectroscopic modes MIRI MRS and NIRSpec IFS, since these are widely used modes of JWST with broad scientific applicability, where the data reduction has been particularly difficult. 

This paper is organized as follows. \S\ref{sec:scicontext} summarizes the scientific context and the philosophy that informed the TEMPLATES program. \S\ref{sec:scigoals} summarizes the science goals, and \S\ref{sec:techgoals} the technical goals. \S\ref{sec:obsprogram} describes the target selection and design of the observations. \S\ref{sec:dataredux} describes at length how we reduced the data, including issues encountered and mitigated. \S\ref{sec:lessonslearned} takes a step back and explores the lessons learned from this early JWST observing program, that should influence the design, execution, and data processing of subsequent programs. \S\ref{sec:deliverables} describes the deliverables that TEMPLATES is releasing, most notably high-level science-ready data products, and Jupyter python notebooks that document exactly how we reduced the data. \S\ref{sec:finalthoughts} closes the paper.

All calculations assume the Planck \citet{Planck.2020}  cosmology unless otherwise indicated.

\section{Scientific and Technical Context}\label{sec:scicontext}

\subsection{Galaxies across the full range of dustiness}

The optical and infrared backgrounds have roughly equal power at wavelengths above and below 3.5~\micron\ \citep{Hauser.2001}, which strongly implies that both unobscured and obscured star formation were important over cosmic history. Spectacular examples of the obscured mode include submillimeter galaxies (SMGs), which include the most luminous, dustiest galaxies known (see review by \citealt{Caseyreview.2014}.) By contrast, UV--bright galaxies selected by dropout techniques such as the Lyman Break Galaxies (LBGs) have low obscuration, and are far more common than submillimeter galaxies, with lower much star formation rates (see review by \citealt{Shapleyreview.2011}.)

Selection of UV--bright galaxies from deep surveys from the \textit{Hubble Space Telescope} has revealed the star formation history of the Universe  \citep{Madau.2014}, a major accomplishment of modern astrophysics.  However, these studies with \textit{Hubble} have in general relied on rest-frame ultraviolet continuum emission to trace star formation --- a diagnostic that is extremely susceptible to attenuation by dust.  Indeed, the most prodigiously star-forming galaxies, which have been found through large--area millimeter and submillimeter surveys like those performed by \textit{Herschel}, \textit{Planck}, and the South Pole Telescope (SPT),  \citep{Negrello.2010, Vieira.2010, Vieira.2013, Everett.2020, Wardlow.2013, Harrington.2016} disappear entirely from \textit{Hubble} surveys due to dust attenuation (e.g., \cite{Walter.2012, Chen.2015, Ma.2015}) 

Though both high-attenuation and low-attenuation galaxies are important to the cosmic history of star formation (\citealt{Zavala.2021}, especially their Fig.~7), the scientists who study each group have historically had almost no data in common, and therefore rarely attend the same conferences.  It's almost like they see different universes.   

We were therefore motivated to study both highly obscured and unobscured galaxies with a common set of diagnostics. JWST is the first observatory that bridges the divide between low-attenuation and high-attenuation galaxies, because it can obtain a common set of spectral diagnostics across the full range of dust attenuation. 
Thanks to tremendous sensitivity \citep{Rigby.2023}, JWST can obtain the polycyclic aromatic hydrocarbon (PAH) dust features in lensed galaxies even when the obscuration is very low.  Such dust diagnostics were extremely difficult for \textit{Spitzer} to obtain spectroscopically in the distant universe \citep{Rigby.2008, Menéndez-Delmestre.2009}.  Likewise, JWST can obtain rest-frame optical spectral diagnostics for $z>3$  galaxies even when the obscuration is very high; such rest-frame optical spectra simply did not exist for highly obscured galaxies before JWST.

Further, JWST can obtain the emission line Paschen $\alpha$ at $\lambda=1.8751~\micron$ (hereafter Pa~$\alpha$) for galaxies with little regard for attenuation. Pa~$\alpha$  is the gold standard diagnostic of star formation rate in the nearby universe \citep{Alonso-Herrero.2006}, as it is extremely robust to attenuation and, as a hydrogen recombination line, it directly measures the recombination rate, and thus (in equilibrium) the ionization rate.  We do note that while Pa~$\alpha$ is the best star formation rate diagnostic available, it may not be perfect; it is possible that Pa~$\alpha$ may be optically thick in the very most obscured galaxies \citep{Simpson.2017, Shuowen.2022}.
While it was extremely difficult for \textit{Spitzer} to measure Pa~$\alpha$ for galaxies in the distant universe \citep{Papovich.2009, Rujopakarn.2012}, with JWST these measurements were expected to be routine.

\subsection{The advantage of gravitational lensing}\label{sec:lensingisnice}
The physical scales corresponding to the diffraction limit of current telescopes do not permit the study of the internal processes of distant galaxies.  Gravitational lensing offers a way to push past this physical limit, to discern important structures such as star-forming regions and star clusters.  Accordingly, numerous programs with \textit{Hubble} and now JWST have targeted strongly--lensed galaxies.

Some worked examples illustrate this point.  The diffraction limit of \textit{Hubble}, 0.034\arcsec\ at $\lambda = 0.4$~\micron, corresponds to a physical scale of 300~pc at $z=1.5$ and 240~pc at $z=4$.  
These are the same physical scales achieved by JWST at 1.1~\micron, the wavelength at which the telescope becomes diffraction--limited \citep{Rigby.2023, McElwain.2023, Lajoie.2023}.  Such unlensed spatial resolution can discern large-scale properties such as bulges and disks, but will blur out all but the largest star-forming regions and stellar clusters.  

For the examples above, because the physical resolution scales as the square root of the strong lensing magnification factor ($\sim\sqrt{\mu}$), a magnification of $\mu = 25$ enables spatial resolution of $\approx$60 and 50~pc at $z=1.5$ and $z=4$, respectively. Thus, gravitational lensing provides the only way to access the internal physical scales that are important for galaxy evolution, like the scales of star clusters, over most of cosmic time.

\subsection{Specific JWST context}
JWST’s incredible sensitivity \citep{Rigby.2023} has revealed galaxies out to very high redshift \citep{Curtis-Lake.2023}, and captured their rest-frame optical and rest-frame UV spectral diagnostics (e.g. \citealt{Bunker.2023,Tang.2023,Fujimoto.2023,Matthee.2023,D'Eugenio.2023}.)   Though JWST's striking images have captured the public's imagination \citep{Pontoppidan.2022}, three-quarters of the general observing time in the first two years has gone to spectroscopy ($70\%$ in Cycle 1\footnote{\url{https://www.stsci.edu/files/live/sites/www/files/home/jwst/science-planning/user-committees/jwst-users-committee/_documents/jstuc-0421-jwst-cycle1-review-package.pdf}}, 
and $77\%$ in Cycle 2\footnote{\url{https://www.stsci.edu/files/live/sites/www/files/home/jwst/science-planning/user-committees/jwst-users-committee/_documents/jstuc-0223-cycle2-submissions-chen.pdf}}).
All four of JWST's science instruments have spectroscopic capabilities, including integral field spectroscopy (IFS) with both NIRSpec and MIRI, multi-object spectroscopy (MOS) with NIRSpec, and slitless spectroscopy with NIRCam and NIRISS.  Spectroscopic programs to drill into the deep fields are measuring star formation rates (SFR), physical conditions, and gas kinematics of galaxies across cosmic time. 

Investigations of distant unlensed galaxies are limited to the spatial scales allowed by the diffraction limit, as discussed in \S\ref{sec:lensingisnice}.  Given these limitations,  the NIRSpec guaranteed time observers chose, for their several hundred hour integral field spectroscopic investigation of 28 unlensed galaxies at redshifts $2 \ge z \ge 6$ (PIDs 1216, 1217, and 1262), to target ``some of the brightest and most extended star forming galaxies and AGN/QSO hosts up to $z\sim7$''\footnote{\url{https://www.stsci.edu/jwst/phase2-public/1216.pdf} and \url{https://www.stsci.edu/jwst/phase2-public/1217.pdf}.}  

Given their ability to push past the diffraction limit, studies of gravitationally lensed galaxies have proven popular for JWST, with 19 approved programs in the first two years of science. Cycle 1 included 13 approved General Observer (GO) programs involving lensing, 4  GTO programs,  %[cite CANUCS and PEARLS, Massimo's GTO PID 1199, and NIRSPEC PID 1262 ]
and 2 approved ERS programs: TEMPLATES (this program) and GLASS \citep{Treu.2022}.
In Cycle 2, 11 lensing--assisted GO programs were approved, including 2 medium programs (PIDs 3743, 4125) and 1 large program (PID 3293).

Given our team's interests in studying galaxy evolution using gravitationally lensed galaxies, we proposed for early release science the program TEMPLATES, a spectroscopy--focused program to study four lensed galaxies that span a large parameter space of attenuation, redshift, and star formation rate.  TEMPLATES not only pioneered lensed galaxy science with JWST, the program also demonstrated highly effective  methods of taking and reducing data from JWST, especially integral field spectroscopy from the NIRSpec and MIRI science instruments.

\subsection{The landscape of early spectroscopic results with JWST} 
In the first few months after the start of JWST science, most of the initial extragalactic science papers used one or both of two science instrument modes:  NIRCam imaging and NIRSpec multiobject spectroscopy (MOS) mode.  Papers using the NIRSpec MOS  mode (for example, \citealt{Curtis-Lake.2023, Bunker.2023, Larson.2023, Carnall.2023}) emerged more quickly than papers using the NIRSpec integral field spectroscopy (IFS) mode, where the pioneers were the ERS program Q3D (PI Wylezalek) \citep{Wylezalek.2022, Vayner.2023, Veilleux.2023}.  In retrospect this trend makes sense: While MOS is the most complex operational mode, its data is simpler to reduce than data from  the IFS modes.  This is especially true for NIRSpec MOS spectroscopy in the default configuration, in which the spectra of spatially adjacent microshutters are subtracted from the targeted microshutter, a process that removes much of the detector's pattern noise.  This detector noise is significant for NIRSpec, and NIRSpec IFS mode has no comparable way to remove it; instead the noise must be corrected at the exposure level.  We believe this detector noise issue has been a main impediment to publishing science results from NIRSpec IFS mode to date; we describe this residual noise and its mitigation in \S\ref{sec:reduxnirspec}.  

For the MIRI MRS mode, cosmic ray showers have been the largest barrier to early publication; we discuss those issues and their mitigation in \S\ref{sec:reduxmirimrs}.  

Mindful of these trends, in this paper we spend considerable effort documenting our data reduction methods for integral field spectroscopy for both NIRSpec and MIRI.  We hope that the community can apply our methods to efficiently obtain high-quality science-ready data from similar  JWST observing programs.

\section{Science Goals}\label{sec:scigoals}

TEMPLATES was built around four science goals:
\begin{enumerate}
\item Demonstrate attenuation-robust SFR diagnostics for distant galaxies.
\item Map the distribution of star formation in distant galaxies.
\item  Compare the young and old stellar populations.
\item Measure the physical conditions of star formation, and their spatial variation.
\end{enumerate}

We now discuss each of these science goals in turn.

\subsection{Demonstrate attenuation-robust SFR diagnostics for distant galaxies} 
What is the relation between obscured and un-obscured star formation? How does one reconcile discrepancies between SFRs measured in the UV and the far-IR?  Does the light of different star formation indicators even come from the same locations in galaxies?  Studies of nearby galaxies like GOALS \citep{GOALS} and SINGS \citep{SINGS} have measured all the major SFR diagnostics: UV continuum, far-IR continuum, H$\alpha$, Pa$ \alpha$, and PAHs. Diagnostics calibrated to these local samples may not we well-suited to the redshifted universe, because we know that galaxies have experienced tremendous evolution in size, star formation rate surface density, star formation efficiency, and gas supply. Unfortunately, high redshift galaxy samples typically have SFR measured either from rest-frame UV continuum or from rest-frame far-IR continuum; rarely do the samples or observables—or indeed research communities —intersect. For instance, vigorously  star-forming ($ SFR >100$~\sfrunit ) galaxies are always accompanied by large amounts of dust attenuation. The inferred SFRs in such galaxies from H$\alpha$  and far-IR can be discrepant by an order-of-magnitude (e.g., \citealt{Takata.2006, Casey.2017}) or even 
two orders of magnitude \citep{Hayward.2018}. 

For Lyman-break galaxies (i.e., those with significant escaping UV emission) it is notoriously difficult to measure dust properties and far-IR luminosities (e.g., Watson \etal\ 2015, Knudsen \etal\ 2017, \citealt{Laporte.2017}). A primary goal of TEMPLATES is to empirically calibrate the SFR estimators in a sample of distant galaxies spanning a broad range of SFR, attenuation, and stellar mass, and to do so on resolved scales corresponding to the individual star forming regions within said galaxies. These calibrations will be used by the JWST user community to inform observations, survey strategies, and interpretation of observables. In addition to using H$\alpha$  and H$\beta$ to measure attenuation-corrected SFR, we make the first spatially-resolved measurements in distant galaxies of Pa~$\alpha$, the gold standard indicator of star formation rate.  

TEMPLATES also spatially resolves the PAH $3.3$~\micron\ line \citep{Siana.2009}, which is observable by JWST out to $z=7$, and is three times brighter than Pa~$\alpha$. Together, these measurements of spatially-resolved, attenuation-robust star formation diagnostics provide common observables across a broad sample of distant galaxies, and lay a foundation for future JWST observations. 

\subsection{Map star formation in distant galaxies} A generic prediction of simulations is that cold gas should accrete onto galactic disks.  Some models predict that at Cosmic Noon ($z$$\sim$$1$), high accretion rates of cold gas should lead to high gas surface densities, resulting in unstable disks that violently fragment into kiloparsec--scale clumps \citep{Genzel.2011, Kereš.2005, Dekel.2006}.  Indeed, HST deep fields have revealed that more than half of $1<z<3$ star-forming galaxies appear to have large (0.5--1 kpc) clumps in the rest-frame UV \citep{Elmegreen.2005z6k,Elmegreen.2007, Elmegreen.2009l8, Shibuya.2016}, which might be evidence for cold flow accretion at work.  Determining the properties of such clumps provides critical tests of these theoretical models. However, these clump sizes are uncomfortably close to the HST diffraction limit. In fact, studies of gravitationally lensed galaxies see no preferred size at 1 kpc, instead resolving star forming regions on spatial scales as small as can be probed, $r \sim 30$--100 pc \citep{Jones.2010, Livermore.2012, Livermore.2015, Johnson.2017, Johnson.2017b, Cava.2018, Dessauges-Zavadsky.2019, Iani.2021, Spilker.2022}. Absent lensing, HST’s normal spatial resolution will blur this highly clumpy star formation into an apparently smooth exponential disk (see Figure 2 of \citealt{Rigby.2017}). Moreover, any highly extincted regions drop out in such data. 	

Star formation in the distant universe looks very different from star formation at low redshift, in several ways.  Intense star-forming regions in cosmic noon galaxies are physically larger than in their low-redshift counterparts, the ultraluminous infrared galaxies \citep{Rujopakarn.2011}.  In the distant universe,  galaxy-wide starbursts may be the more typical mode than the nuclear starbursts that dominate in the local universe \citep{Gladders.2013}.  The mid-infrared spectra of luminous galaxies at cosmic noon match those of $z \sim 0$ galaxies of much lower luminosity \citep{Rigby.2008}, rather than luminosity--matched counterparts --- suggesting that intense star formation at early times may have lower optical depths and be more spatially extended.  The ISM pressure is significantly higher in high redshift galaxies than in local star-forming galaxies, which facilitates the formation of H$_2$, allowing molecular clouds to cool and collapse (e.g., Popping \etal\ 2014). While giant molecular clouds in the Milky Way and local galaxies follow well--known scaling relations, such as that between the cloud size and linewidth, the limited observations of cloud-like structures at high redshift indicate that they lie significantly above the local relations \citep{Swinbank.2015, Dessauges-Zavadsky.2019, Spilker.2022}, likely as a consequence of the increased ISM pressure. Direct, detailed studies of star formation at high redshift are extremely important and relevant for models of galaxy evolution; local analogs do not suffice.  

To produce the extreme SFRs observed in high redshift submillimeter galaxies, theoretical models have appealed to a diverse range of processes, such as gas-rich major mergers (e.g., \citealt{Chakrabarti.2008, Narayanan.2010}), violent disk instabilities (e.g., \citealt{Ceverino.2015,Lacey.2016}), significant gas infall from the IGM \citealt{Narayanan.2015, Lovell.2021} or hybrid processes \citep{Hayward.2011, Hayward.2012, Hayward.2013}. Even then, producing a realistic population of submillimeter galaxies is challenging, as models cannot simultaneously match the observed number counts of submillimeter galaxies while simultaneously matching that of massive quenched galaxies at the same epoch \citep{Hayward.2021}.

From spatially-resolved observations of H$\alpha$, H$\beta$, and Pa~$\alpha$, TEMPLATES will measure the morphology, clump luminosity distribution, and clump size distribution of attenuation-corrected star formation in four highly magnified galaxies. For typical lensing magnifications, JWST's near-infrared instruments resolve star-forming regions with sizes smaller than $100$~pc. By measuring the sizes and luminosities of star-forming clumps as small as 30 Doradus and Carina, TEMPLATES will characterize the importance of clumps, and enable spatially-resolved measurements of the physical conditions of star-forming regions.

This science goal requires mapping the attenuation inside galaxies, which prior to the JWST era, had not been done for field galaxies in the distant universe except by stacking at 0.5 kpc resolution \citep{Nelson.2016}.  Attenuation is starting to be mapped using JWST slitless spectroscopy and the Balmer decrement on spatial scales down to 0.3~kpc for field galaxies  \citep{Matharu.2023}, with larger samples coming. The only way to measure attenuation on smaller spatial scales for distant galaxies is with strongly lensed galaxies (e.g.~\citealt{Patricio.2019, Claeyssens.2022}.)

\subsection{Compare the young and old stellar populations.} 
A key result from SDSS and \textit{Spitzer} is the so-called ``star formation main sequence'': that a galaxy’s stellar mass predicts its total SFR \citep{Brinchmann.2004, Noeske.2007}.  TEMPLATES will, for the first time at these redshifts, obtain attenuation-robust specific star formation rates ( $sSFR \equiv\  SFR / M_*$ ) for both LBGs and SMGs, thus placing them in the SFR–$M_*$ plane, and contextualizing their stages of galaxy evolution. Due to extreme dust obscuration, stellar masses of submillimeter galaxies could not be reliably measured with pre-JWST facilities (Michałowski \etal\ 2012; Ma \etal\ 2015). Furthermore, SFRs are typically measured heterogeneously, preventing direct, robust comparisons between various star formation observables, and calling into question the very concept of a main sequence of star formation. 

By resolving the sSFR relation in these galaxies, the TEMPLATES data will shed light on the origin of these scaling relations that link a galaxy's past and present.  Comparing the spatial distribution of star formation rate and stellar mass will show how star formation progresses spatially over time—whether galaxies form from the inside-out or the outside-in \citep{Sánchez-Blázquez.2007, Perez.2013}.
%HST grism H$\alpha$ spectra \citep{Nelson.2016}point to inside-out growth, but that work is subject to systematics that are unavoidable with HST, namely that hundreds of galaxies had to be stacked to build signal, attenuation is not well measured, and there is an unknown  amount of contamination of H$\alpha$ by [N II]. By making these measurements on individual galaxies, with control of dust attenuation, and in a highly spatially resolved way, we will measure the impact of these systematics.

\subsection{Measure the physical conditions of star formation, and their spatial variation.} 
We set the NIRSpec integration times to ensure adequate SNR in H$\beta$ and H$\alpha$  for individual regions within the target galaxies. Such integrations are sufficiently deep to also obtain the full suite of diagnostics from rest-frame 0.44~\micron\ to 0.80~\micron, for two of the four TEMPLATES targets. These diagnostics measure the metallicity, ionization parameter, and pressure of the nebular regions of these galaxies, through comparisons to photoionization and shock models (e.g. \citealt{Kewley.2013, Kewley.2019, Sutherland.2017}). The TEMPLATES spatially resolved JWST spectra also measure how much these diagnostics vary spatially within each galaxy. This enables an estimate of the extent to which gradients may bias the integrated-galaxy measurements made by Keck and VLT for thousands of star-forming galaxies at these redshifts (e.g. \citealt{Sanders.2016, Strom.2017}); stacking of HST grism spectra \citep{Trump.2011} and now JWST NIRISS grism spectra \citet{Matharu.2023} indicate the effect may be significant.

For the Lyman break galaxies these observations will improve upon the spatial resolution available from the ground. Due to the extreme dust content in submillimeter galaxies, optical spectroscopy has been notoriously difficult and biased towards galaxies with unobscured sight-lines at lower redshift (e.g. \citealt{Swinbank.2004, Takata.2006, Casey.2014, Casey.2017, Danielson.2017}). Basic measurements, like metallicity and reddening, had to wait for TEMPLATES. In addition to providing the first robust, unbiased optical spectroscopic study of submillimeter galaxies, TEMPLATES provides the first spatially-resolved, rest-frame optical spectra in dusty, luminous galaxies.

In addition, by comparing the rest-frame optical emission line diagnostics to the mid-IR continuum and PAH strengths, we can quantify any contribution from AGN to the observed energy output.   The mid-IR continuum, which is emitted by hot dust grains around the central engine of a supermassive black hole is one of the most distinctive ways of identifying active galactic nuclei (AGN), including those that are heavily obscured by dust.

For the two TEMPLATES targets that are submillimeter galaxies (SPT0418$-$47 and SPT2147$-$50), the NIRSpec, MIRI/MRS, and NIRCam data from TEMPLATES were partially analysed by \citet{Birkin.2023, Cathey.2023, Spilker:2023}. H$\alpha$ and the [N{\sc ii}] doublet are detected at high S/N in the integrated NIRSpec spectra; the [N{\sc ii}]/H$\alpha$ ratio, and thus the metallicity, are spatially resolved for these galaxies. Both sources show apparently near-solar metallicities, and SPT2147$-$50 in particular displays regions where [N{\sc ii}]/H$\alpha$ is greater than unity, which is interpreted as evidence for previously undetected AGN emission. Previous analysis of SPT0418$-$47 suggested that it was a kinematically cold rotating disk \citep{Rizzo.2020}, but the improved spatial resolution from NIRCam allowed \citet{Cathey.2023} to identify an interacting companion at a separation of 4.4\,kpc, with a mass ratio of approximately 4 to 1. The 3.3\,$\mu$m PAH feature was also detected in this source by the MIRI/MRS \citep{Spilker:2023}, currently the most distant and only spatially-resolved PAH detection at high redshift. The MIRI data suggest that SPT0418$-$47 does not obviously host an obscured AGN, and that large spatial variations in the ratio of PAH to IR luminosity make this PAH feature a complicated tracer of star formation (at best).

\subsection{Ancillary Science}
The JWST data and rich ancillary datasets enable additional science beyond these four science goals.  Some of these include:  comparison of  the dust mass and gas mass as revealed by ALMA with the current star formation as revealed by JWST; comparison of the galactic outflows seen in H$\alpha$ by JWST with outflows seen in rest-frame UV spectra from Keck and Magellan and molecular absorption from ALMA; determination of whether outflows depend on star formation surface density;  searches for dwarf galaxies and dark matter sub-structure in the foreground lenses;  determination of the AGN contribution to the mid-IR emission from submillimeter galaxies; and studies of the lenses themselves (two galaxy clusters and two early-type galaxies).

We look forward to and encourage papers using TEMPLATES data that will be written by the community beyond our team, enabled by the high-level data products we are providing.

\section{Technical Goals}\label{sec:techgoals}
In addition to scientific merit, the ERS programs were chosen to obtain representative datasets early in the lifecycle of the JWST mission, to obtain information that would help the community prepare observing  proposals, and to engage a broad cross-section of astronomers and planetary scientists.  As such, TEMPLATES had two technical and community--oriented goals, which we now discuss.

\subsection{Optimize the JWST spectroscopic pipeline}  
Based on our experience with the early days of the \textit{Spitzer} mission, we anticipated that generating science-ready spectroscopic data cubes would be the most challenging aspect of this program.  We expected, at the least, to have to tune the parameters in the pipeline that control background subtraction, removal of fringing and stray light from MRS data, removal of the MSA imprints from NIRSpec IFS data, and outlier detection.   In addition, before launch it was not at all clear how impactful would be the optional IFS calibration frames (dedicated off-source backgrounds for both MIRI and NIRSpec, and MSA leak calibration frames for NIRSpec).  Therefore, TEMPLATES set the technical goal of shaking out and optimizing the JWST spectroscopic pipeline for both MIRI MRS and NIRSpec IFS.  We have done so, and describe our best practices in \S\ref{sec:dataredux}.

\subsection{Establish best practices for integral field unit spectroscopy with JWST}
Following the best practices described by JDox  before launch  \citep{JDOX}, our observations included off-source background observations for both MIRI and NIRSpec, as well as NIRSpec leak calibration files which were intended  to correct light leaking through the closed microshutter array onto the detector.  Before launch, it was not at all clear whether dedicated background observations are required, or whether it would be possible to derive the background from the periphery of on-source frames. Similarly, it was not clear before launch how necessary were MSA leak calibrations; we therefore obtained MSA leak calibrations for every dither position, with the idea that we could test whether on-source dithering alone, or dithering plus a smaller number of leak cals, could sufficiently correct IFS data for leaks from stuck open MSA shutters and print--through.  Our plan was to determine best practices for these types of observations, such that future IFS observations could be streamlined to the extent possible.  \S\ref{sec:nirspec_strategy} presents our proposed best practices for observing galaxies with NIRSpec IFS mode.

\section{Observing Program Design}\label{sec:obsprogram}

\begin{figure*}
\includegraphics[scale=0.58]{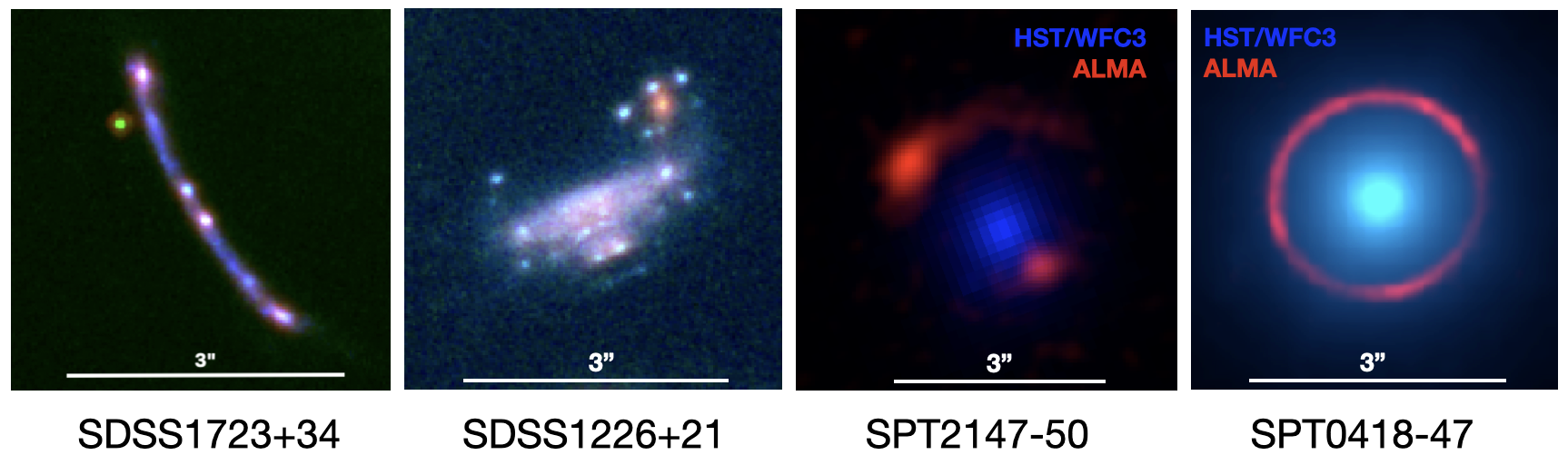}
\figcaption{The TEMPLATES sample. The HST filters used were F390W, F775W, and F110W for SGAS1723$+$34; F606W, F814W, and F110W for  SGAS1226$+$21; and  F140W for the two submillimeter galaxies. The ALMA band is rest-frame 160~\micron\ continuum.  A 3\arcsec\ scalebar is shown to illustrate the approximate size of the NIRSpec IFU field of view.  
The TEMPLATES galaxies are all highly magnified, with sizes that fit in the fields of view of the JWST IFUs. The sources are all aligned with North up and East left.
\label{fig:samplefigure}}
\end{figure*}

\subsection{Target selection}\label{sec:targets}
The four TEMPLATES targets (Table~\ref{tab:targets} and Figure~\ref{fig:familyportrait}) have been extensively studied \citep{Kubo.2010, Koester.2010, Bayliss.2011, Wuyts.2012, Weiss.2013, Vieira.2013, Spilker.2014, Stark.2013, Rigby.2015,  Gullberg.2015, Aravena.2016, Ma.2015, Spilker.2016, Rigby.2017, Chisholm.2017, Rigby.2018,  Chisholm.2019, Sharon.2020, Spilker.2020.paperI, Florian.2021, Rigby.2021, Gururajan.2022, Solimano.2022}.

We selected two submillimeter galaxies from the South Pole Telescope (SPT, \citealt{Vieira.2010, Ma.2015, Spilker.2016}), although we considered all far-IR-selected lensed systems (e.g., Herschel and Planck-- selected). To be considered, we required: a) a spectroscopic redshift; b) high resolution imaging with ALMA and HST; c) a robust lens model; d) Einstein radius small enough ($<1.5$\arcsec ) to fit inside the NIRSpec IFU FOV; e) high spatial resolution resolved spectroscopy of molecular lines with ALMA; and f) visibility within the ERS window. The two selected targets are SPT0418$-$47 and SPT2147$-$50.

Table~\ref{tab:sampleprops} lists the properties of these targets. The table shows that these four targets span a range of luminosity, star formation rate, stellar mass, and attenuation.  All targets have published lens models \citep{Spilker.2016, Spilker.2020.paperI, Sharon.2022},
and extensive imaging and spectroscopic ancillary data (including HST, ALMA, Keck, and Magellan).

\begin{figure*}
\includegraphics[scale=0.25]{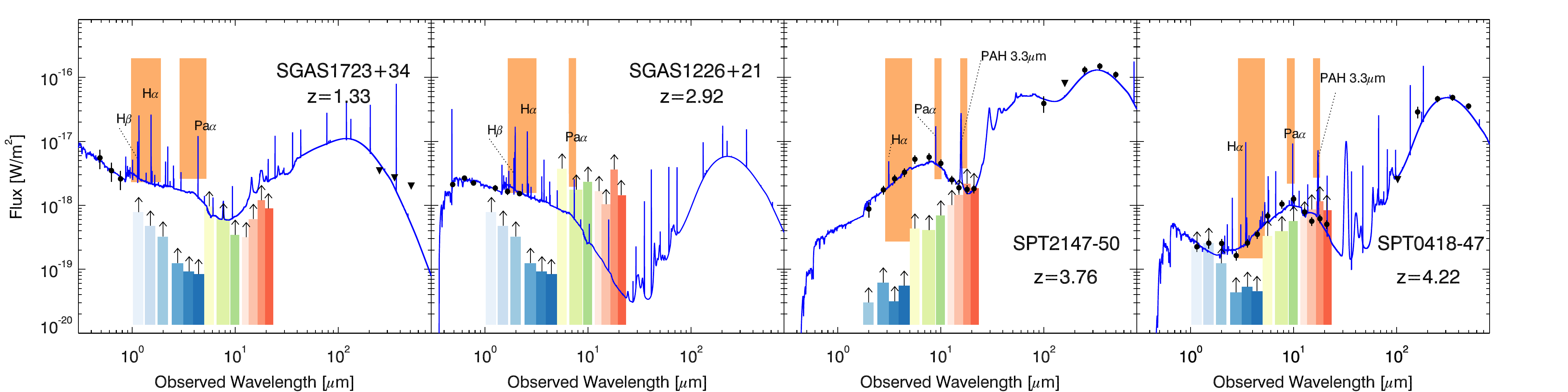}
\figcaption{ Expected intrinsic (demagnified) spectral energy distributions for all the sources with the proposed 5${\sigma}$ depths for imaging shown as upwards arrow from each filter band. 1${\sigma}$ depths for IFU spectroscopy shown in orange. The depths were calculated assuming a point source using the JWST ETC v1.1.1%
\label{fig:sed4x1}}
\end{figure*}

\subsection{Observations}\label{sec:observations}
Table~\ref{tab:obslog} lists the observations that comprise the TEMPLATES program.  In all, there were 53.7~hr of successful observations, plus 10 hours of observations that failed due to a bug in the ground system, and were re-observed.  Here, we summarize the strategy for each observing mode. Further details can be found in the APT file for our program, which can be retrieved using APT from STScI by querying for program ID 1355.

\subsubsection{Imaging}
We designed imaging observations for TEMPLATES using  the NIRCam \citep{Rieke.2023} and MIRI \citep{Wright.2023} instruments, covering  4--6 NIRCam filters and 7 MIRI filters per target, to efficiently (in terms of observing time) obtain photometry sufficient to measure stellar mass, constrain the presence of AGN, and map the PAH features.  For the two Lyman break galaxies, observed fluxes from SDSS were used to estimate integration time, conservatively assuming even flux distribution; depths were adjusted to achieve $S/N >30$ per spatial resolution element.  For many filters, this was achieved in the shortest practical integration time.  For the submillimeter galaxies, CIGALE SED fits were used to estimate integration time, and the depth set to $S/N >10$ per resolution element.  For both the Lyman break galaxies and submillimeter galaxies we relax the SNR thresholds for the few cases where the integration per filter would exceed 10~min.

For the NIRCam observations, we selected filters for each target that span a broad wavelength range and would enable robust, resolved SED fitting when used in combination with MIRI imaging.  The original plan was to use 6 filters for all targets (F115W, F150W, F200W, F277W, F356W and F444W); however, after receiving the data for SPT0418$-$47, we decided, given the low SNR in the short wavelength filters, that it would be better to instead go deeper in F200W, dropping the F115W and F150W observations. Total exposure times ranged from 290s to 2750s for targets with 6 filter data; for SPT2147$-$50 the F200W total exposure time was 5840s. The associated total execution times can be found in Table \ref{tab:obslog}. Observations were taken using the INTRASCA dither type with 2-3 primary dithers. Observations were taken with readout patterns BRIGHT 1/2 or SHALLOW2 to avoid saturation of nearby galaxies. 

All four targets were imaged by MIRI in seven imaging filters (F560W, F770W, F1000W, F1280W, F1500W, F1800W, and F2100W). The total execution times for all filters are listed in Table~\ref{tab:obslog}.   Across the four targets and seven filters our exposure times vary from 1--14 minutes. All four targets were supposed to use a dither pattern optimized for point source, as the sources were small in size compared to the Imaging Field of View (FoV). On examining the MIRI imaging data for the first two targets  (SGAS1226$+$21, SGAS1723$+$34), we noticed an implementation issue with the dither pattern, where the source was not centered in the imaging field of view, and was instead closer to the edge by the last filter in our observation sequence which was the longest wavelength filter(F2100W). Our program instrument scientist for MIRI suggested we change the dither pattern to be optimized for extended sources for the remaining two targets (SPT0418$-$47 and SPT2147$-$50), as the fix for the dither implementation issue was not going to be in place before these targets were observed.

\subsubsection{NIRSpec spectroscopy}
For each of the four galaxies in our sample, we used the NIRSpec integral field unit (IFU) \citep{Böker.2022}, which provides spatially-resolved imaging spectroscopy for a 3\arcsec\ $\times$ 3\arcsec\ field of view, with 
$0.1 \times 0.1\arcsec$ spatial sampling.  Each galaxy had IFS observations taken with one grating to target their rest-optical/NIR light.  The lowest redshift galaxy in the sample, SGAS1723$+$34, was observed with a second NIRSpec grating setting to cover Pa~$\alpha$.

The NIRSpec IFS observations for each source, including grating/filter pair, exposure times, etc., are summarized in Table\ref{tab:obslog}.  For the two LBGs, we used the high--resolution gratings (R$\sim$2700) to achieve the highest spectral resolution possible with this instrument, to resolve kinematics within these sources.  For the two SMGs, we used the medium resolution gratings (R$\sim$1000) to achieve higher throughput for these fainter sources. This was a conservative choice, given that the attenuation of these objects was not well-known prior to JWST.

NIRSpec IFS  observations were taken with the NIRSIRS2 readout pattern \citep{Rauscher.2017}, as our sources are faint, and used small--size cycling dithers. Following pre-launch guidance from JDox, we obtained dedicated off-source background pointings and leak calibrations for all sources at each dither position. 

The initial NIRSpec observations of SGAS1723$+$34 did not complete successfully. The telescope pointing drifted, causing the target to fall outside the IFU field of view. This issue was quickly noticed, and the full set of SGAS1723$+$34 NIRSpec observations were scheduled to be retaken (WOPR 88493). However, we found that the first set of exposures taken with the G140H grating on the science target completed successfully before the drift began. We therefore ended up with twice the exposure time in this grating. We include this extra set of G140H in our final data reduction.

\subsubsection{MIRI MRS spectroscopy}
TEMPLATES  uses the MIRI Medium Resolution Spectroscopy (MRS) integral-field observing mode for three galaxies, primarily targeting Pa~$\alpha$ for all targets except the lowest--redshift source (SGAS1723$+$34, for which NIRSpec rather than MIRI captured Pa~$\alpha$), and 3.3$\mu$m PAH emission for both submillimeter galaxies.  Given the faint expected fluxes, We did not attempt to capture the PAH emission in MRS for the two Lyman break galaxies.

Given the expected observed redshifts of the targeted emission lines, SGAS1226$+$21 was observed using only the LONG (`C') MRS grating, while SPT0418$-$47 and SPT2147$-$50 were observed using both the LONG and MEDIUM (`B') gratings; see Table~\ref{tab:obslog}.  Because the MRS observes four disjoint wavelength ranges in each grating setting, several other spectral lines such as Brackett $\alpha$, Brackett $\beta$, Paschen $\beta$ and molecular hydrogen rotation lines fall in the observed bandpass for each galaxy, although these features are typically expected to be fainter than the primary lines of interest.

All MIRI MRS observations used the SLOWR1 detector readout pattern and a standard four-point extended-source dither pattern. Total on-source exposure times varied between about 40 and 60~min, depending on the predicted flux of the targeted line. The MRS observations were accompanied by four-point dedicated background observations for an equal integration time. The MIRI imager in the F560W or F1000W filter was used during the background exposures. For SPT2147$-$50, we realized that we could use target offset coordinates to ensure that the source was covered by the imager during the MRS background exposures, resulting in very deep F560W and F1000W imaging ($\sim$2400 and 2800s, compared to $\sim$100s in the dedicated imaging exposures). We did the same for SGAS1226$+$21 MIRI MRS observations and obtained a deep F560W image ($\sim$1 hr). Although the dither pattern is not optimized to sample the imager PSF, the extra depth will prove useful in the future to verify the astrometric registration and provide very deep observations of the target sources.

Since the MRS fields of view are not fully concentric at all wavelengths it was not possible to simultaneously center the targets in every spectral channel.  We adopted a mixture of centroiding optimized for all four channels (``primary channel'' set to ``ALL") and optimized for individual channels containing the spectral line of interest.  Ultimately, due to the effects of cosmic ray shower artifacts (see Sec.~\ref{sec:reduxmirimrs}), it would have been more advantageous to always center the targets in the respective target channels, because our shower removal technique relies on having source-free areas on all sides of the source.

\section{Data Reduction and Calibration}\label{sec:dataredux}

\subsection{Downloading raw data}
We downloaded the processed raw data (level 1b) from MAST.  The TEMPLATES github site publishes a simple script, adapted from one written by Richard Shaw of STScI, that downloads any given dataset given the proposal ID and type of data desired. We have found this script much easier to use than the JWST MAST interface that was available for the first 1.5 years of the JWST science mission.   

\subsection{MIRI imaging}\label{sec:reduxmiriimaging}

Given TEMPLATES' focus on spectroscopy, the MIRI imaging was shallow compared to other JWST ERS programs. Initial inspection of level 3 products from MAST portal showed vertical striping patterns in all the filters and targets, indicating the presence of detector $1/f$ noise. We experimented with several of the striping methods being used by the community to determine the best way to de-stripe our data. Uncalibrated images were downloaded from the MAST portal and processed through the \jdrp\ calibration pipeline \citep{Bushouse.2022} version 1.11.3 using calibration reference data system pipeline mapping (CRDS, pmap) 1106. % TAH updated pipeline notation
We implemented stripe removal for the imaging data by removing column and row trends \footnote{following notebook https://github.com/STScI-MIRI/Imaging\_ExampleNB/blob/main/helpers/miri\_clean.py} 
after stage 2 of the pipeline (‘*cal.fits files’). This de-striped stage 2 data was processed through the stage 3 pipeline, and the images after stage 3 were used for all analysis.

Apart from the striping issue, we also came across an issue with persistence arising from a strong cosmic ray hit that is not currently addressed by the automatic pipeline. A cosmic ray hit during the F770W observation of  SGAS1226$+$21 at an oblique angle. This was flagged correctly in the pipeline for the processing of F770W images, but  created a non-linear response in the same region of the detector for the observations of subsequent filters (F1000W through F2100W). The cosmic ray was not identified in the subsequent filters because it did not produce a `jump' in the count rate ramps during those later filters; it was present throughout the exposure. To mitigate this issue, we changed the data quality flag in stage 1 of the observations for filters F1000W through F2100W, using a ds9 region file in  detector coordinates.  Dithering helped get acceptable level 3 products for this source despite this issue.  The function to implement this fix is available on our team’s github.

\subsection{NIRCam imaging}\label{sec:reduxnircam}

\begin{figure*}
\centering
\includegraphics[width=\columnwidth]{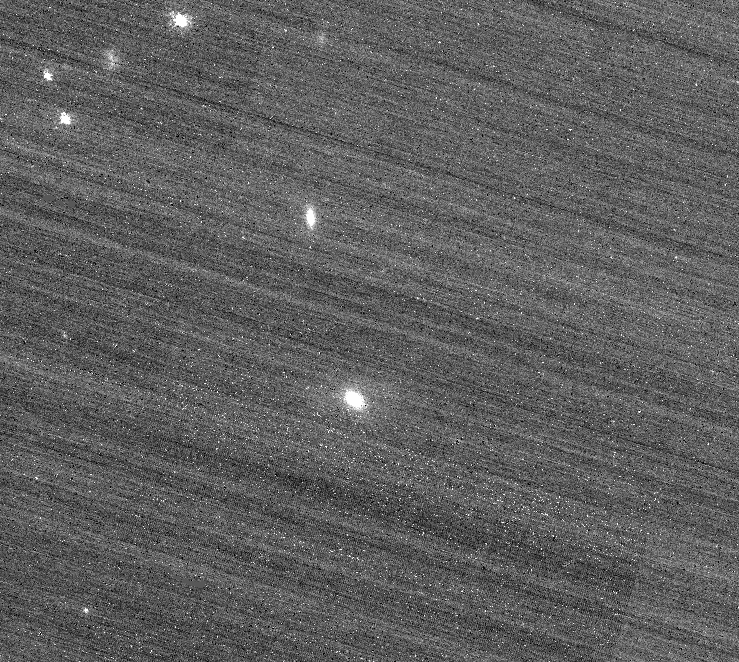}\includegraphics[width=\columnwidth]{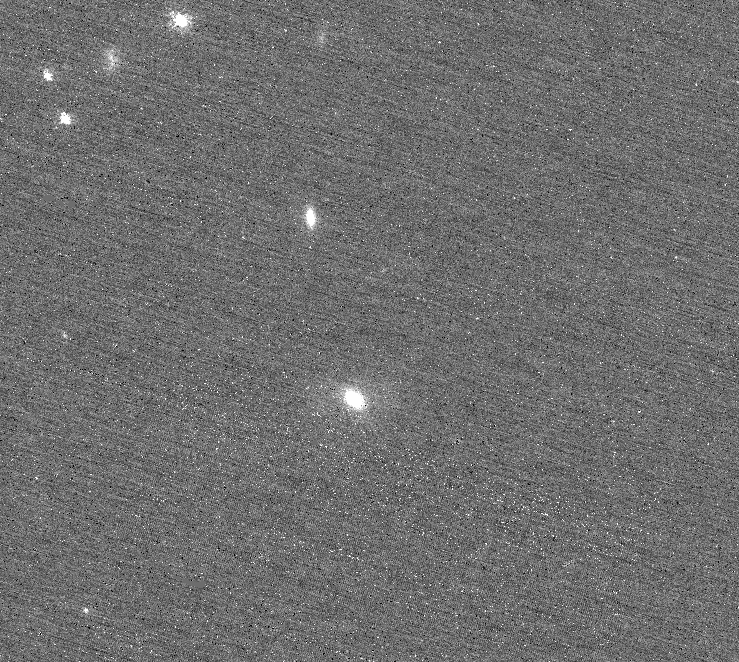}
\figcaption{JWST NIRCam imaging of SGAS1723$+$34 before (left) and after (right) applying our custom destriping algorithm. %[XX State which direction rows, columns are running.  State where the bright target is, and approx FOV.]
\label{fig:destipe}}
\end{figure*}

\begin{figure*}
\includegraphics[scale=0.45]{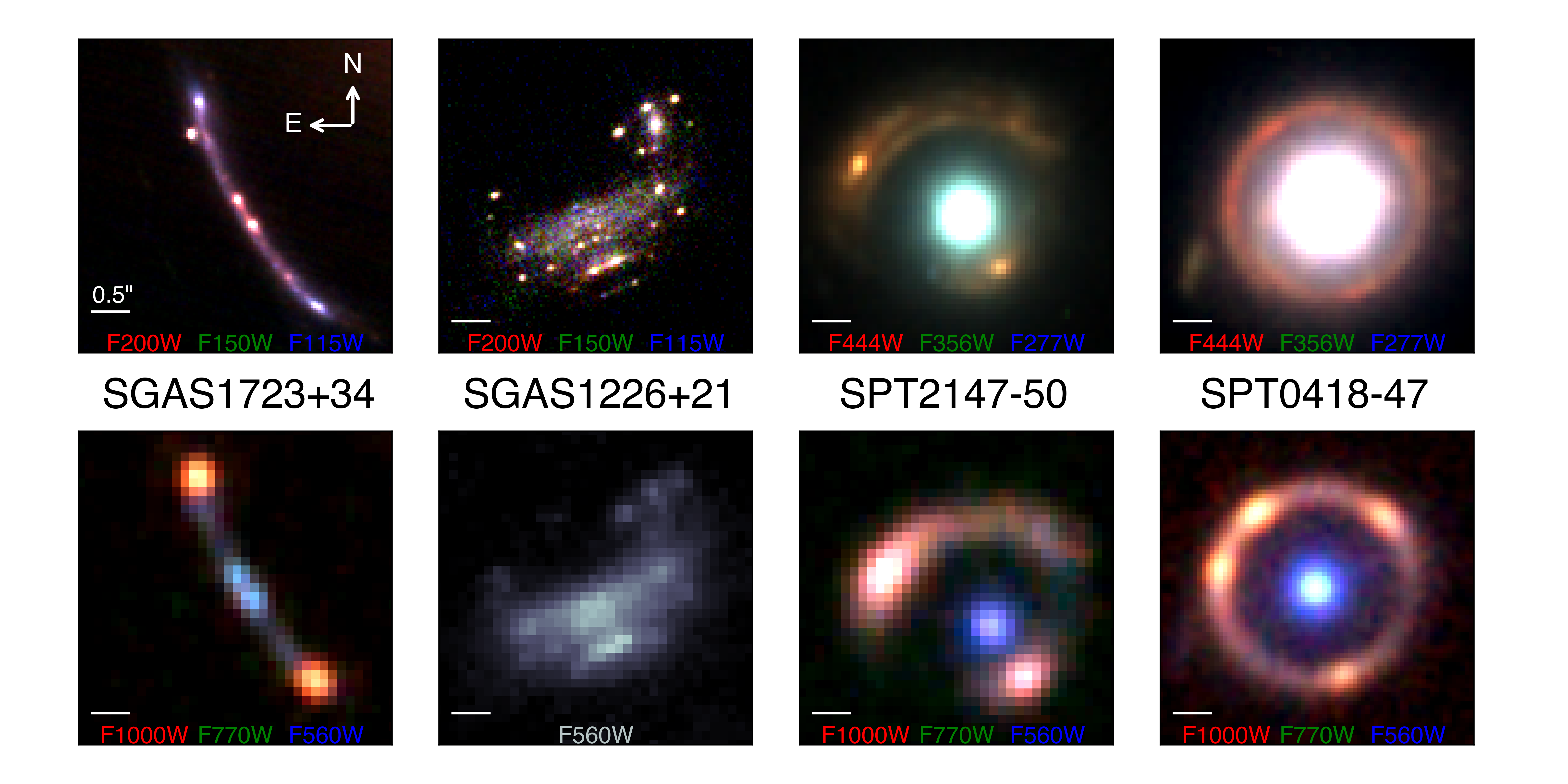}
\figcaption{JWST NIRCam (top row) and MIRI (bottom) row imaging for the TEMPLATES program. NIRCam short wavelength filters F200W, F150W and F115W were used for the two Lyman break galaxy sources, whereas long wavelength filters F444W, F356W and F277W were used for the two submillimeter galaxies as the red/green/blue (RGB) channels. For MIRI imaging, F1000W, F770W, and F560W were selected as the filters for RGB for all sources but one.  The exception, SGAS1226$+$21, is only detected in F560W, so only that filter image is shown using the colormap ``bone''. All images have been aligned with North up and East left, with a common scale bar of half an arcsecond shown.
\label{fig:familyportrait}}
\end{figure*}

For the NIRCam imaging, we started with the Level 2a data products. The first step in our processing work flow is to perform a custom de-striping of the Level 2a data, as the $1/f$ detector noise is significant for these short exposures. This procedure also corrects for residual amplitude offsets between different amplifiers in the detector.  The de-striping proceeded as follows: 1) an object mask of the individual frame was created by thresholding an initial version of the processed and stacked science image, which was then propagated back to the individual frame level assuming the initial astrometric solution; 2) all unmasked pixels were used to compute and subtract a median value for each of the four amplifier regions, and then 3) each masked pedestal-corrected amplifier region was filtered using a horizontal median filter 512$\times$1 pixels in extent, the result of which was then subtracted.
Figure \ref{fig:destipe} shows single exposure images of SGAS1723$+$34 before and after the destriping procedure.

The destriped images were then processed using the standard \jdrp\ pipeline (version 1.11.0). We have posted in the TEMPLATES github repository the parameter files used to reduce each source. The NIRCam Jupyter notebooks we publish show how to generate these parameter files.

%In order to control the many different parameters between reductions we have been generating parameter files, these can be generated by running the individual reduction pipeline through the command prompt with the condition --save-parameters {filename.asdf}. These parameter files may have to be regenerated after pipeline updates as more parameters are added to the pipeline stages. 
Most parameters for NIRCam reduction did not need tweaking; however, initially we found it necessary to turn off alignment to GAIA in the pipeline, as that led to considerable astrometric offsets between filters. With newer pipeline and reference file updates, astrometric alignment using GAIA DR3 as absolute reference catalog shows astrometric registration at sub-pixel level in the NIRCam LW filters. However, NIRCam SW still has issues with astrometric registration \footnote{https://github.com/spacetelescope/jwst/issues/7993}. To overcome this issue we created a catalog with GAIA DR3 registered shortest wavelength filter of NIRCam LW (i.e. F277W) and used it as the absolute reference catalog in the tweakreg step of stage 3 in the imaging pipeline. We found it necessary to do this correction for each module separately and hope that future reference file updates will fix this issue.
%{\color{red} Add a cautionary note here about versions and proper astrometry? Jared, do you want to include a figure showing the astrometric offsets between filters when using both the Gaia stars and internal alignment?} -- I need to re-check if that's still an issue with the current pipeline versions, but we could with the qualifier that it's from older pipeline versions. I haven't made one since we decided to turn off GAIA alignment. 
Ongoing updates to the calibration data have also led to increasingly unified photometry between instruments and improved consistency with HST. %(reference Grace's work?). %(and telescopes?).%should we reference the Science Performance Paper and/or include any of Grace's work with JWST/Hubble bands for photometry?
We defined the photometric calibration files that were released October 3, 2022 to be the standard, and have either updated the photom values for earlier reductions, or re-reduced the data.
Finally, we used WebbPSF \citep{Perrin.2014} to generate models PSFs based on wavefront sensing data from before and after each of our NIRCam observations.  The JWST wavefront is measured roughly every two days \citep{Rigby.2023}. In modeling galaxy morphology (Cathey et al.\ submitted to ApJ) using GALFIT \citep{Peng.2006}, we found no appreciable difference as to whether the ``before'' or ``after'' PSF was used; this makes sense given the excellent stability of the JWST PSF \citep{McElwain.2023, Lajoie.2023}.

\subsection{NIRSpec spectroscopy}\label{sec:reduxnirspec}

We reduced the NIRSpec IFS data using the \jdrp\ calibration pipeline version 1.11.3, using CRDS pmap 1105.  The pipeline processes the data in the following order: 1) Stage 1, detector-level corrections and ramp-fitting made to the individual raw data products from the instrument; 2) Stage 2, instrument mode-specific calibrations, including WCS and wavelength solutions, flagging of failed-open MSA shutters, flat fielding, path loss correction, and flux calibration, made to individual exposures; 3) Stage 3, data combined from multiple exposures for a given observation, resampled onto a common grid and coadded into a single data cube.

The \jdrp\ pipeline offers many customization options, and we outline our parameter choices here. 
For Stage 1, pipeline versions 1.9.6 and later include an \texttt{expand\_large\_events} option, which is designed to expand large jump detections to eliminate ``snowball'' artifacts. 
This step is turned off by default, but here we choose to include it to better mask snowballs. 
The Stage 2 pipeline is nominally where dedicated leakcals would be subtracted from the data. However, we choose not to subtract the leakcals taken as part of this observing program. 
The observed fields do not contain many bright stars, so fairly little light leaks through the MSA onto the detectors.
Including leakcals therefore only add detector noise, so we choose not to use them.

Finally, for Stage 3, we %choose to turn off the \texttt{outlier\_detection} step, following advice from the STScI helpdesk.  This step of the \jdrp\ pipeline does not yet appear to work well.  The way that the step tends to fail for TEMPLATES data is to flag the bright emission lines for rejection. 
use a two-pronged approach in removing outliers from the data.  First, we use the updated \texttt{outlier\_detection} step in the pipeline (in \jdrp\ versions 1.11.3 and later) to remove the majority of the outliers present.  However, additional outliers remain after this step that require further processing.  %To replace this pipeline step, 
We developed a layered sigma-clipping routine to post-process the final reduced data cubes and remove the remaining outliers.   In brief, this routine sigma clips the off-galaxy spaxels in a uniform manner, then takes the galaxy spaxels and clips them in layers, separated into 3--4 bins set by the signal-to-noise (S/N) ratio of the brightest line in the spectrum.  We briefly describe this routine in \S\ref{sec:sigmaclip} and further detail the algorithm and code release in Hutchison et al. (submitted to PASP).

The NIRSpec pipeline cube building step involves a 3D drizzling step that combines multiple dithers into a single data cube \citep{Law.2023}. The standard cube building step produces a final cube with 0\farcs1 square spaxels, however the spaxel size is tunable through the \texttt{scalexy} parameter. Many of the TEMPLATES sources have an abundance of substructure visible in the high-resolution imaging; some of this structure is not well resolved with the undersampled 0\farcs1 native spaxel size. We therefore additionally produced cubes with 0\farcs05 pixels. We find that these higher spatial resolution data cubes better resolve small structures in our lensed arcs without introducing additional artifacts. 

%[More here about each stage, what options we choose.  Don’t forget to mention whether leak-cals were subtracted, and how/whether the background was subtracted.
%outline - to be sentence-ified later - 
%Stage 1 - expand\_large\_events turned on to mask snowballs.
%Stage 2 - No leak cals subtracted - they just add noise, and we're in pretty dark fields so there's not much leakage.
%Stage 3 - outlier\_detection turned off due to early issues, replaced with layered sigma clipping described below.
%Background subtraction is done after the stage 3 pipeline. We use the JWST backgrounds tool to calculate the expected total background for our observations. Then we create a cube where each spaxel is the expected backround, in MJy/sr (same output unit as s3d files). That background cube is then subtracted from the Stage 3 output to create our background-subtracted products.]

\subsubsection{Residual pattern noise from NIRSpec data}
The NIRSpec data show correlated pattern noise along columns.  This noise is caused by milli-Kelvin temperature fluctuations in the SIDECAR ASIC chips that control and read out the NIRSpec detectors (B.~Rauscher, priv.\ comm.)  The resulting residual pattern noise is not removed by the IRS$^2$ noise-reduction readout mode \citep{Rauscher.2017}.  In our data, the root-mean-square of this noise is about 2~e$^-$ in each of the two detectors.  This pattern noise changes on short timescales, such that one exposure cannot be used to correct the noise in the next exposure, because the noise has changed too much.  As a result, dedicated background observations cannot remove this noise; instead, it must be removed at the exposure level.

This noise is most problematic for IFS mode, since there can be no subtraction of spectra from nearby rows, as is the case for fixed-slit mode or for MOS mode when a source is nodded up and down among several shutters.  However, it has also proven beneficial to correct MSA data for this pattern noise  (\citealt{Strom.2023} J.\ Chisholm priv.\ comm.)

In the spirit of developing best practices for observing and data reduction, in the next subsection we describe in detail how we corrected for this noise, including our tests of algorithms developed by three different groups.

\subsubsection{Removing residual pattern noise: The great NIRSpec bakeoff}\label{sec:bakeoff}

We tested three different methods of removing the residual  NIRSpec pattern noise, and visually compared both the 2D countrate images and the final 1D spectra to determine which method works best.
We wanted to test the methods on both NIRSpec detectors, NRS2 and NRS1, since they have different noise properties; NRS2 is noisier.  This required examining high-resolution spectra, which cover both detectors, rather than medium resolution spectra which span only NRS1.  We therefore chose the SGAS1723$+$34 data for these experiments, since that TEMPLATES target has NIRSpec observations taken with two different high-resolution gratings.   The pattern noise is most prominent in the G140H grating, since the zodiacal background is relatively low at those blue wavelengths \citep{Rigby.2023bg}.   We applied the  pattern noise removal methods after the Stage 1 pipeline is run, on the output countrate \texttt{*rate.fits} files. 

%[State which stage of the data the methods are applied to.  In other words, where is is inserted into the pipeline flow? - after Stage 1, before Stage 2]

The first noise removal method we tested was the “basic median” approach, developed by Stephan Birkmann of the NIRSpec instrument development team, and provided to TEMPLATES by STScI's JWST helpdesk. This method calculates a column-by-column median, then subtracts that median from each pixel in the column. To avoid removing signal, this method utilizes only the dark pixels located between the IFU traces and the fixed slit region. This region is covered by the support structure around the NIRSpec slits, and is thus blind to astrophysical light. However, relatively few rows are utilized compared to the full size of the detector.  %[Say who we got this idea from] 

The second strategy we tested was a ``rolling median'' approach, developed by Ian Wong and provided to TEMPLATES through private communication. This method applies a median filter of length 201 pixels to each column. The median filter allows for variation across the width of the detector, which should provide a more accurate representation of the detector background than the basic median approach. This method requires that the detector backgrounds be sampled across the full span of the detector face, and thus must first mask out all illuminated pixels. The mask is created using a combination of the data quality flags from the \texttt{rate} files (to mask out cosmic ray ``snowballs'') and the spectral traces in the Stage 2 pipeline output calibrated \texttt{cal.fits} files, which mark all non-illuminated pixels as NaN. The fixed slits must be masked manually, as they are not captured in the \texttt{cal} files. The median filter length of 201 pixels was chosen to be long enough to span the largest masked sections of the detector, and thus produce a smooth background to be subtracted. 

The third and final strategy we tested is the NSClean software, described by \citet{Rauscher.2023}.  This method models the detector noise in Fourier space for each pixel column. Like the rolling median, this method samples the full width of the detector, providing a robust estimate of the instrument backgrounds. 

Properly masking the illuminated pixels is critical for the success of the NSClean method. We use a similar masking technique as described above for the rolling median method, in which we remove IFU spectral traces using the \texttt{cal} files, and manually remove fixed slits, to produce a mask that blocks out all illuminated pixels. Masking ``snowball'' artifacts caused by cosmic ray impacts is also important. We tested two methods to remove these artifacts prior to running NSClean - removing every flagged jump detection from the pipeline-produced data quality array, and manually removing large snowballs identified by eye. 
We found that masking all pixels with a flagged jump detection caused additional artifacts to appear in the \texttt{rate} files after applying NSClean, likely from large sections of detector pixel columns being masked leading to underconstrained fitting. Therefore, when running NSClean, we do not mask snowballs using the data quality flags, but instead visually inspect each \texttt{rate} file for prominent snowballs prior to fitting. Any large snowballs found are masked manually.

\begin{figure*}
    \centering
    \includegraphics[width=0.9\textwidth]{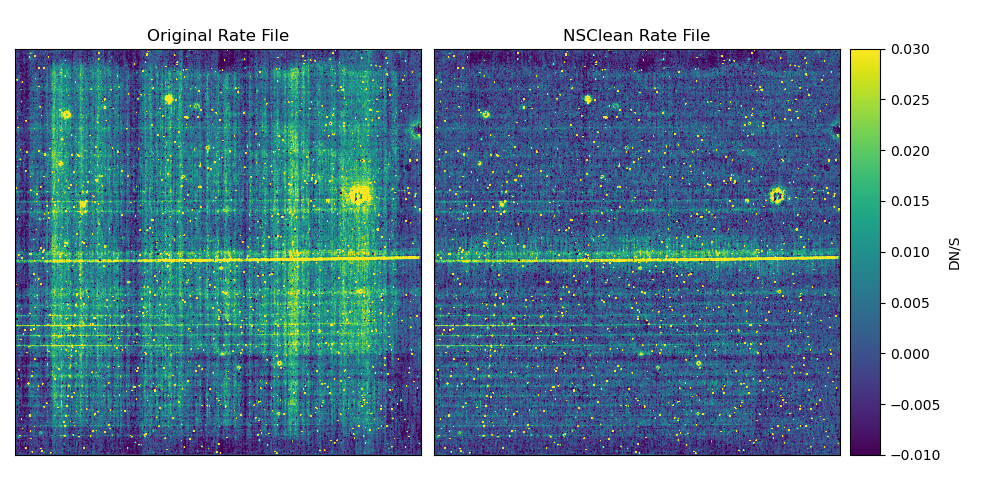}
    \caption{2D detector images (rate.fits files) before and after NSClean. The vertical stripes present in the pre-NSClean image create artificial wiggles in the extracted spectra (see Fig. \ref{fig:nsclean_1D}). These vertical stripes are effectively removed by the algorithm, enabling better measurements of the continuum shape of the final spectrum.}
    \label{fig:nsclean_2D}
\end{figure*}

We visually compared the cleaned \texttt{rate} files that emerged from each of the three noise reduction methods; see  Fig.~\ref{fig:nsclean_2D}. 
Each method provides marked improvement over the original \texttt{rate} images, offering some correction to both the overall detector bias and the vertical banding. Of the three methods, NSClean best removes the pattern noise.  In second place is the rolling median, which also produces an acceptable result.  The basic median approach leaves the most residual detector noise of our three tested strategies. 

\begin{figure*}
\centering

\includegraphics[width=0.495\textwidth]{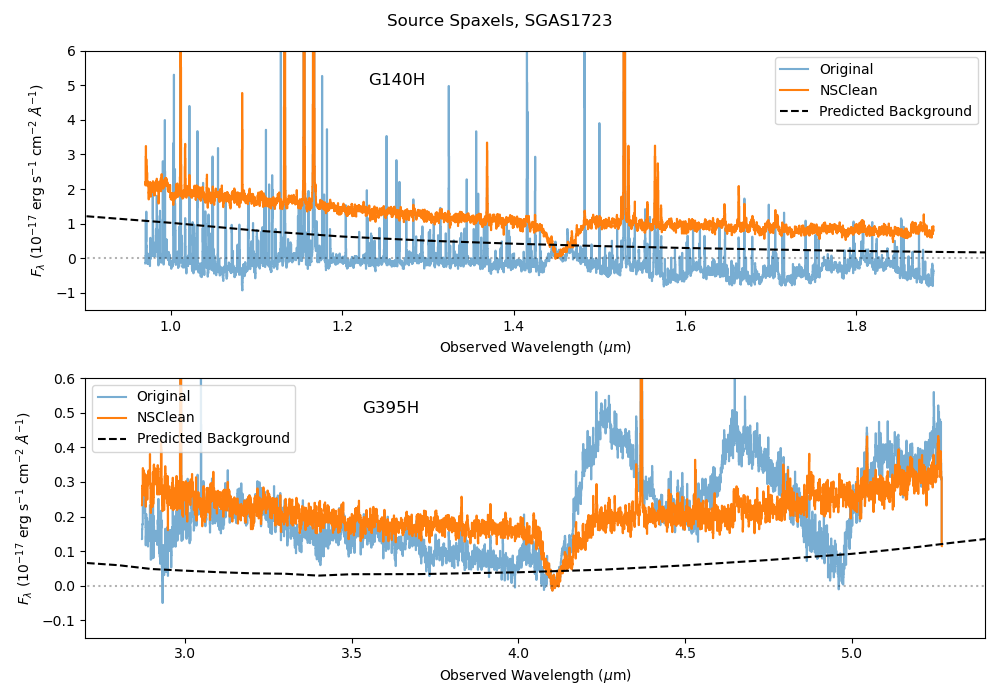}
\includegraphics[width=0.495\textwidth]{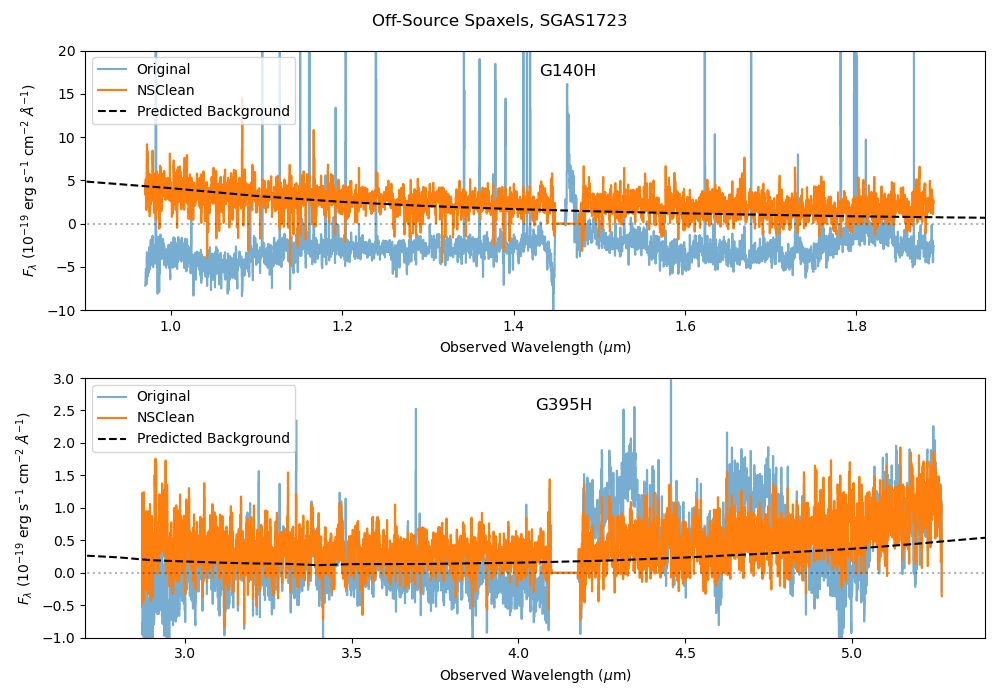}
\caption{Extracted 1D spectra are shown in the left two panels, using a custom aperture where included spaxels have SNR $> 3$ in the bright [OIII]$\lambda5008$ emission line. The upper left panel shows the on-source spectrum from the G140H grating, while the lower left panel shows the on-source spectrum from the G395H grating. The right hand panels show the off-source spectra for each grating, created using the inverse of the source aperture. In each panel, blue lines show the extracted spectra before applying any $1/f$ noise correction, while orange lines show spectra after applying NSClean. Black dashed lines show the expected background calculated from the JWST Backgrounds Tool. The original spectra show both fluctuations and overall offsets in the continuum level caused by the $1/f$ noise in the detectors, while these noise features are removed by NSClean. }
\label{fig:nsclean_1D}
\end{figure*}

Ultimately, we want to know how each of these solutions translates to a final reduced spectrum. We therefore ran each cleaned set of rate.fits files through the Stage 2 and Stage 3 pipelines, following the same standard procedure for each set (i.e., not subtracting leak calibrations or background exposures). Visual inspection of the resulting spectra lead to the same conclusion as the rate files, namely that NSClean provides the cleanest end product (Fig. \ref{fig:nsclean_1D}). The basic median cleans the spectra considerably relative to the initial product, however some residual correlated noise features remain. The rolling median approach removes the vertical striping in the rate files (and thus the "wiggles" in the final spectra). However we found it to be a less reliable method than NSClean to remove overall offsets in the continuum level (e.g. the original negative continuum flux seen in the blue spectrum in Figure \ref{fig:nsclean_1D}). We therefore conclude that NSClean provides the best result of the three methods.  We therefore apply NSClean to all our NIRSpec IFS observations, and recommend its use for other programs (as discussed in \S\ref{sec:lessonslearned}.)

\subsubsection{Custom Outlier Rejection of the NIRSpec Data Cubes}\label{sec:sigmaclip}

For the first year of JWST science operations, the outlier detection and rejection step of the \jdrp\ pipeline, which works on the individual dithers, was not working correctly for NIRSpec IFS data. Initially this was due to issues with the NIRSpec astrometric solution, such that the algorithm was comparing source brightness for pixels that should sample the same sky position in multiple dithers, but in fact were not.  For the remainder of the first year, the main issue was overzealous behavior by the algorithm, such that real, valid strong emission lines were being flagged and removed from our data.  In July 2023, an update dramatically improved the performance of the \jdrp\ pipeline's outlier rejection (versions 1.11.3 and later).  

In order to be able to work with the TEMPLATES data for the first year, our team developed a custom method of outlier rejection (Hutchison et al., submitted), which works on the drizzled data cubes \citep{Law.2023}.
Now that the pipeline's default outlier rejection is working much better, this custom approach is less essential.  That said, we find that using the pipeline's updated outlier rejection, and then a final clean-up of the drizzled data cubes using our custom method, produces better results than does the pipeline alone. Hutchison et al.\ (submitted) describe our algorithm,  and quantitatively compare the effectiveness of the two outlier detection approaches for TEMPLATES NIRSpec IFS data.  We release the outlier rejection code itself at \href{https://github.com/aibhleog/baryon-sweep}{github.com/aibhleog/baryon-sweep} (DOI: 10.5281/zenodo.8377532).

We release final outlier-rejected data cubes as deliverable; see \S\ref{sec:deliverables}. We also release the masks that separate target spaxels from the rest of the cubes, which are used by our custom outlier rejection code. The creation and use of these masks are described in Hutchison et al.\ (submitted).

\subsection{MIRI medium resolution spectroscopy}\label{sec:reduxmirimrs}

Our basic data reduction workflow is described in more detail in \citet{Spilker:2023}, including our implementation of a method to remove the so-called ``cosmic ray shower" artifacts \citep{Argyriou.2023}. In brief, we generally followed the default \jdrp\ pipeline procedures as of version 1.8.4, with a few additional processing steps. After the \texttt{Detector1} pipeline was run, we used the dedicated background exposures to identify and flag additional bad pixels not already masked by the pipeline bad pixel map. In the \texttt{Spec2} pipeline, we performed a 2-D pixel-by-pixel background subtraction by median combining the individual background exposures. This allows for the removal of detector and flat-field systematic effects that otherwise limit our ability to recover faint, extended emission. We ran the \texttt{Spec3} pipeline in its default configuration to produce a 3-D data cube for each of the four MRS channels in each observed grating setting \citep{Law.2023}.

\begin{figure*}
\includegraphics[scale=0.5]{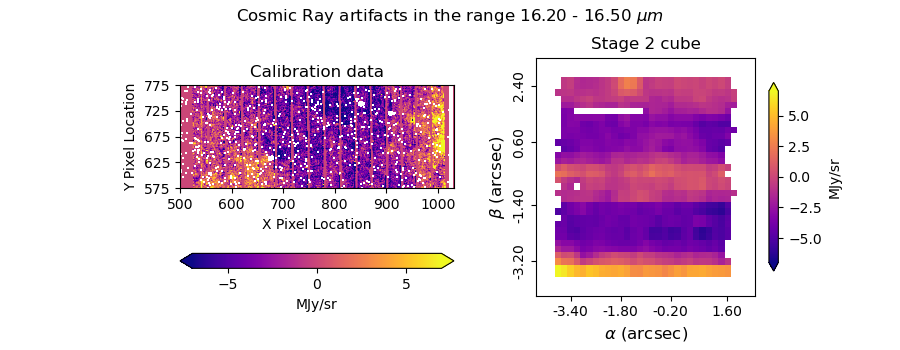}
\figcaption{Demonstration of cosmic ray showers detector artifacts after being processed through stage 2 of the pipeline. Left: 2-D detector image after stage 2 in one dither position. There are no bright emission lines expected in the region selected. We can see the detector artifacts varying in flux values where we expect a smooth background. Right: Collapsed cube for the same region shown in the left hand side of the figure. The detector artifacts manifest as stripes in the 3-D cube.
\label{fig:showers}}
\end{figure*}

By far the dominant instrumental systematic, that remains in the data after pipeline processing, are residual unflagged pixels resulting from the cosmic ray showers.  These artifacts, illustrated in Figure~\ref{fig:showers}, limit the sensitivity of the data. 
%Showers persist despite our use of a \jdrp\ pipeline version with a preliminary treatment of these artifacts. 
Showers are present in both the on- and off-source exposures, resulting in both positive and negative artifacts in the final data cubes. 
While a preliminary treatment of these artifacts in the \jdrp\ pipeline has significantly improved these artifacts for some science programs, this treatment does not currently work well for SLOW-mode data such as ours.
Due to the geometry of the showers and the MRS slicing optics, the showers result in stripes in the 3-D cubes that are mostly aligned with the cube's x-dimension. As described in detail in \citet{Spilker:2023}, we removed these artifacts by estimating the level of the stripes using a series of rows in the cube x-dimension after masking regions with real source emission and subtracting this `stripe template' from the cube. 

Our current data processing produced MRS cubes that are science-ready \citep[e.g.][]{Spilker:2023}, but we are continuing to investigate improved data reduction methods. Among these include alternative methods to identify and remove the effects of showers, which we expect will be an active area of research in the coming years. 
We also continue to investigate the best method of obtaining and using dedicated background exposures for MRS science data.

\section{Lessons Learned}\label{sec:lessonslearned}

\subsection{Use the latest versions of pipeline and calibration products}
A basic lesson that we learned, is that users should assiduously keep up-to-date both the \jdrp\ pipeline code and the calibration files; these changed frequently, with considerable impact on the results of data reduction, during the first 1.5 years of JWST science operations.  We also noted that odd results (such as flux densities changing by several orders of magnitude) can occur if an older pipeline version is run with the latest calibration products.  It also seems important, after updating either the pipeline or the calibration files, to re-run all steps of the pipeline, rather than ``mixing and matching'' versions.

To keep the pipeline up to date, one should frequently use the command \texttt{pip install --upgrade jwst}.   To keep the calibration files up-to-date, one should set the environment variable \texttt{CRDS\_SERVER\_URL} to point to the  CRDS webpage\footnote{\href{https://jwst-crds.stsci.edu/}{https://jwst-crds.stsci.edu/}}.  This will cause the calibration files to automatically update each time the pipeline is run, assuming the computer is connected to the internet.  The TEMPLATES NIRSpec reduction notebook demonstrates how to set this environmental variable.

\subsection{ \texorpdfstring{$1/f$}{1/f} noise in short imaging exposures}
TEMPLATES spent most of its observing time on spectroscopy.  Imaging exposure times were designed to reach signal-to-noise goals given the expected range of spectral energy distributions.  As such, given the fantastic sensitivity of JWST \citep{Rigby.2023}, these integration times were short, only a few hundred seconds per NIRCam filter.  These short imaging times also helped TEMPLATES fit into the $\sim 50$~hr average size of ERS programs.  However, such short integration times are especially subject to $1/f$ noise, which required custom destriping routines to mitigate.  In retrospect, such extremely short integration times may not have been a good idea.  {\bf We therefore recommend} that users consider requesting more than the bare minimum of imaging integration time needed for bright extragalactic targets, since longer integration times make the data easier to reduce, as well as support ancillary science.

\subsection{Correcting residual detector noise in NIRSpec spectroscopy}
As detailed in \S\ref{sec:bakeoff}, the NIRSpec detector shows pattern noise, measured at a level of 
$\sim 2$~e$^-$ rms in our data, which if not corrected, will dominate the noise in integral field mode, and may contribute substantially to noise in the multiobject and fixed slit modes.   In IFS mode, if not corrected, this pattern noise alters --- substantially so, for our targets --- the flux density and shape of the extracted continuum.  {\bf We therefore recommend} that users apply the NSClean algorithm \citep{Rauscher.2023} to correct this noise.  Indeed, STScI is in the process of adding NSClean as a standard step in the NIRSpec pipeline, based in part on our demonstration of its effectiveness for the TEMPLATES data.

\subsection{Strategies for NIRSpec integral field spectroscopy mode }\label{sec:nirspec_strategy}
When planning the TEMPLATES program, from the information available before launch, we could not determine by how much dedicated background observations and leak calibrations would improve the quality of the data.  In the spirit of Early Release Science, we therefore decided to take these calibrations, measure their effect, and then recommend to other users whether these calibrations were in fact needed.

When a dedicated ``background” observation is used in spectroscopy, it is usually intended to subtract one or both of two very different effects: a) the real astrophysical background on the sky, and b) residual detector noise. For NIRSpec, it was not understood before launch which, if either of these effects might require dedicated backgrounds, so the JDox documentation recommended taking these calibrations.  Below, we explain why we believe that dedicated backgrounds are not needed for NIRSpec IFS observations like TEMPLATES'. 

It is worth a brief digression to remind the reader that for NIRSpec, the relative contributions of astrophysical background and detector background will depend on the filter and grating.  NIRSpec prism mode will be dominated by poisson noise from the background sky emission (a combination of zodiacal light, Galactic emission, and scattered light, see \citealt{Rigby.2023bg}), not detector readout noise.  By contrast,  for the medium and high spectral resolution gratings, detector noise (aka read noise) will dominate; this is especially obvious in the blue high-resolution modes, where the zodiacal background levels are relatively low.  As such, for the medium and high resolution gratings used in TEMPLATES, it is far more important to address the residual detector noise, than to precisely subtract out the astrophysical backgrounds.

On-orbit experience shows that dedicated background observations are \emph{not} required to remove the astrophysical backgrounds from NIRSpec data for sparse extragalactic fields, for two reasons.  First, the infrared astrophysical backgrounds are sufficiently well-known \citep{Kelsall.1998}, and the scattered light, stray light, and self-emission properties of the observatory are sufficiently well-understood \citep{Rigby.2023bg}, that the background spectrum can be reliably predicted using the JWST Background tool\footnote{\url{https://jwst-docs.stsci.edu/jwst-other-tools/jwst-backgrounds-tool}}.  This statement is true for wavelengths $\lambda > 1.2$~\micron; the background tool is currently incorrectly extrapolating blueward of COBE's short--wavelength cutoff at $1.2$~\micron.  Second, out of the plane of the galaxy, the astrophysical backgrounds are generally dominated by zodiacal light, which is smooth on arcsecond scales; thus, so long as the target does not completely fill the small ($\sim 3$\arcsec) field of view of the NIRSpec IFU, the background levels can be measured from the periphery.  Thus, for extragalactic targets that do not completely fill the NIRSpec IFU field of view, dedicated off-target background observations need not be obtained. 

The other justification for taking dedicated backgrounds would be to remove residual detector noise.  However, the on-orbit reality is that the NIRSpec pattern noise changes completely from exposure to exposure, since the pattern noise is caused by small rapid fluctuations in the temperature of the readout and control chips \citep{Rauscher.2023}.  Thus, dedicated backgrounds are utterly useless for removing the residual detector noise, because 
the noise in the science exposures will be completely different from the noise in the background exposures.   Instead, the noise must be corrected within each exposure, as we demonstrate in \S\ref{sec:bakeoff}.

{\bf We therefore recommend that users \emph{not} take dedicated background observations when using NIRSpec with the medium or high resolution gratings longward of $1.2$~\micron} for extragalactic targets.  Only for the low resolution (prism) mode may it make sense, since these observations will be dominated by background noise; or when the spectral shape at $\lambda < 1.2$~\micron\ is important to the science goals.  Even in these two cases, dedicated backgrounds may not be necessary since the background levels for $\lambda > 1.2$~\micron\ can be predicted using the JWST background tool, and since the background levels can be measured from the periphery of the data cube, assuming the source does not fill the entire IFU.

\subsection{Effect of showers on MIRI MRS spectroscopy}
One unexpected result of cosmic ray hits on the MIRI detectors has been the so-called ``shower'' artifacts. Current understanding suggests these showers arise as charge diffuses outward from the location of cosmic ray hits, but unlike the near-IR detectors, in MIRI these showers are typically not circular (or elliptical). Because the counts from the showers are much lower than typical cosmic ray hits and can persist for long periods of time (in some cases even between integrations), the showers themselves are often not automatically identified and flagged by the pipeline's outlier detection strategies. Showers are long-duration events, and in severe cases can persist through a detector reset into the next exposure.

The TEMPLATES MRS data is severely impacted by these showers, which limit our ability to reach the full expected depth of the observations suggested by the ETC and other tools. In principle the effect of these showers should be able to be mitigated by using shorter integration times, since the number of pixels impacted by showers increases in longer exposures and only uncommonly severe showers persist through a detector reset. A larger number of short exposures, however, must be balanced against the read noise penalty. While TEMPLATES used a standard 4-point dither pattern with individual integration times ranging from 600-900\,s, shorter integration times would likely have resulted in lessened shower impacts. Before launch, STScI asked us to switch the MRS observations to the \texttt{SLOWR1} detector readout pattern to lower data volume. In hindsight, a faster readout pattern would have allowed a more precise time sampling (and potential for flagging and removal) of showers, which would have decreased the fraction of data affected, although this is likely to be a small fraction of the total on-source time for most programs.
%{\bf We therefore recommend that one way to mitigate showers is to avoid using the SLOWR1 detector readout pattern for MIRI MRS, if possible.}

Although the current pipeline contains a preliminary algorithm that attempts to identify and flag cosmic ray showers by approximating them as elliptical regions following strong cosmic ray hits, we found that residual effects of the showers were still the limiting factor in our attempts to detect very faint spectral features. TEMPLATES has developed an algorithm, described more fully in \citet{Spilker:2023}, to fit for and remove these showers in the 3-D data cube space. Due to the typical size of the showers on the detectors and the geometry of the slicing optics, the showers appear as horizontal stripes in the IFU-aligned cubes. The stripes can be both positive and negative, since showers in the dedicated background exposures are subtracted from the on-source data.

Briefly, we mask the region of the cube where real source emission was expected based on pre-existing ALMA data. We then fit for a `stripe template' consisting of horizontal rows in the cube using a 25-channel running average in the spectral dimension. We allow for a linear slope from the left to the right half of each row, since the slicing optics are not perfectly aligned with the 2D detector coordinates and the showers have complex morphologies. The resulting cube after subtraction of the stripe template preserves the source flux but removes the strong striping artifacts. This technique only works if sufficient source-free pixels exist \textit{on either side} of the source location. As such, it would have been better if the TEMPLATES sources had been centered in the field-of-view of the MRS channel of interest instead of a position optimized for the field-of-view of all four channels.

Clearly the best way to address the shower artifacts is to mitigate them before data are taken. Unfortunately shower mitigation, which drives toward a larger number of short exposures, conflicts with overall noise considerations pushing for few long exposures. For the TEMPLATES data specifically, a larger number of shorter integrations would likely have resulted in higher-quality data, but we stress that this may not be the case for all science programs. Future investigations into the statistics and prevalence of showers will be needed before concrete recommendations can be made. {\bf We recommend that MRS users carefully consider the tradeoffs between cosmic ray shower mitigation (favoring more short integrations) and overall detector noise (favoring few long integrations). We also recommend that the source be placed at the center of the aperture of the channel of primary interest } (assuming only a single channel is of primary interest), to allow for a more robust estimate of the shower-induced striping in the cube from source-free pixels. If showers are present, our technique is able to remove the shower artifacts for cases where every bit of signal-to-noise is needed. We hope that continued improvements to the pipeline software will one day render our technique unnecessary.

\subsection{Discovery of a bug affecting target groups}
Our observation 7, which executed on 2022 July 15 UT, was corrupted by an until-then-undiscovered bug in the ground system affecting the APT feature known as a target group, which allows a given observing sequence to be repeated for multiple targets within a target group.  
% It was in OPGS, but I don't think the readers will care, so generalizing to "ground system"
The bug was not found in commissioning, since to the best of our knowledge target groups were not used in commissioning. The reason for target groups is to conserve use of the science instrument's moving mechanisms, since they have limited lifetimes.  Due to the bug, instead of chopping back and forth between the source and the off-source position, as the observation worked through the grating/filter combinations, the observation instead dithered further and further away from the source.  Webb Operation Problem Report (WOPR) number 88493 was filed, and the observation was rescheduled as observations 27 and 28, with a workaround of using separate source and sky targets to avoid the target group bug.  The ground system was patched to fix this bug on 2022 Sept.~8 (T.~Keyes and K.~Peterson, priv.~comm.)
%OPGS subsystem implementation. The fix was provided as a patch under PPS version 14.19.7 and was formally 5implemented in the ground system on 8 September 2022
The other NIRSpec IFS observations in this program also used the workaround, so they have separate observations for target and sky.
{\bf Since target groups now work, there is no longer a need for a recommendation on this issue.}

\section{Deliverables}\label{sec:deliverables}

TEMPLATES delivers two main kinds of products to the science user community.  

\begin{figure*}
\centering
\includegraphics[width=0.95\textwidth]{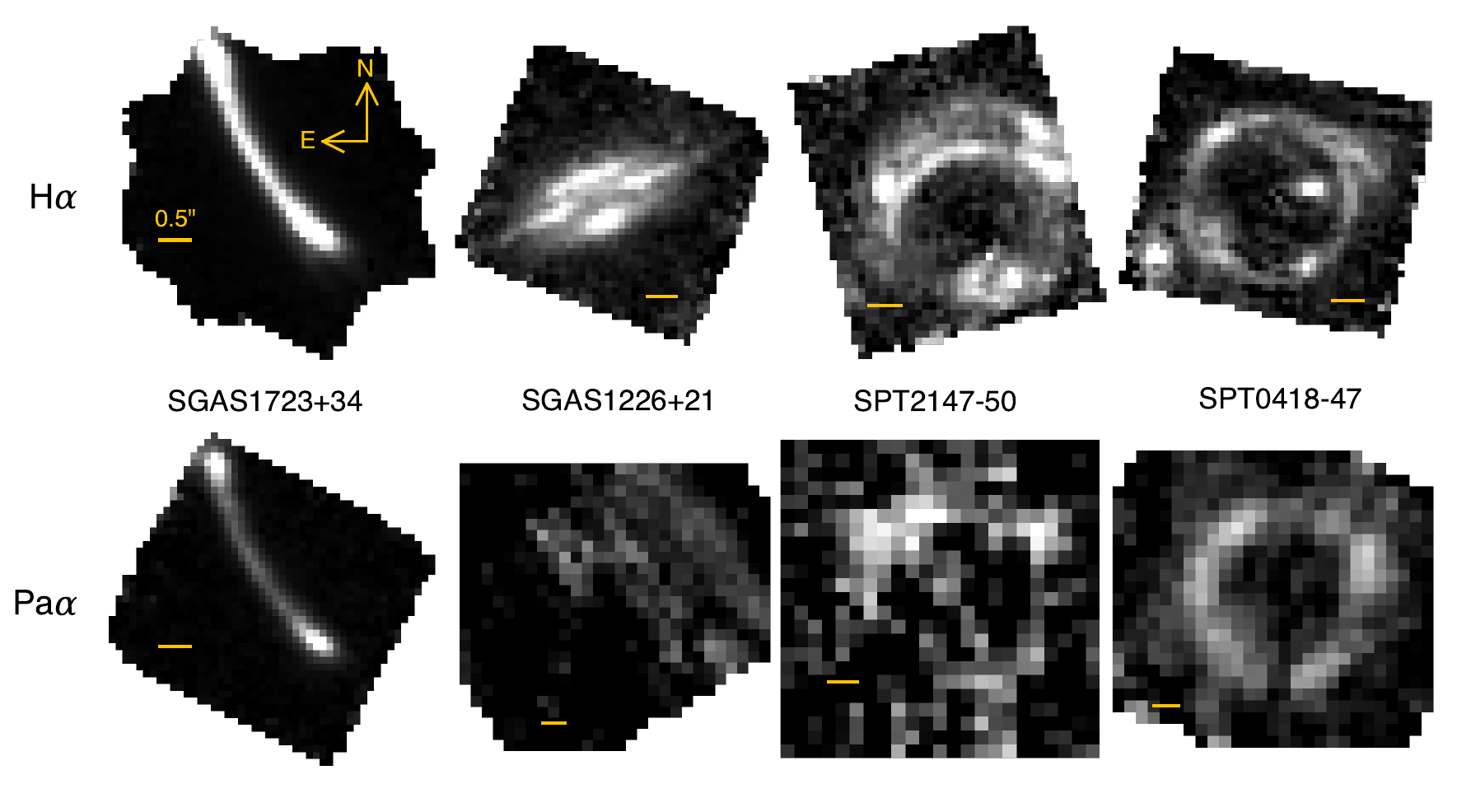}
\figcaption{JWST NIRSpec and MIRI integral field spectroscopy for the four TEMPLATES targets, showing spatially-resolved emission line maps for continuum-subtracted H$\alpha$ and Pa$\alpha$.  The top row and the first panel in the bottom row are medium- and high-resolution NIRSpec integral field spectroscopy, while the final three panels in the bottom row are MIRI Ch1 and Ch2 medium resolution spectroscopy.  All maps have been aligned with North up and East left, with a common scale bar of half an arcsecond shown.
\label{fig:momentzero}}
\end{figure*}

\subsection{Cookbooks} 

Concurrent with the publication of this overview paper, TEMPLATES is delivering how-to cookbooks, that show exactly how we reduced all of our data into science-ready form.  These are Jupyter python notebooks that use the \jdrp\ pipeline as well as our custom software.  Our intent is that other users with similar data, especially MIRI MRS and NIRSpec IFS data, can simply follow our step-by-step notebooks to efficiently produce science-ready data products.  Feedback from a small number of beta testers has been positive.

Our notebooks are available at \url{https://github.com/JWST-Templates/Notebooks}.  We request that researchers follow the citation guidelines in each notebook.

Also in the notebook repository, we release a step-by-step guide to reducing the NIRSpec IFS data from our program, that is intended to be digestible by researchers of any level of experience. This guide covers all relevant steps beginning with installing the JWST data reduction pipelines, through to producing generating sigma-clipped NIRSpec IFS datacubes.  

\subsection{Science-ready JWST data products}
TEMPLATES will release high-level science-ready data products such as fully reduced images and reduced spectral data cubes, as well as  derived data products such as attenuation maps and lens models.  These products will be released on our MAST page at \url{https://archive.stsci.edu/hlsp/templates} as science papers using these products are published by our team.

\subsection{Supporting data from \textit{Hubble}, \textit{Spitzer}, and ALMA}
High-level science products from \textit{Hubble}, \textit{Spitzer}, and ALMA were delivered to STScI in July 2022 and released on MAST\footnote{\url{https://archive.stsci.edu/hlsp/templates/}} as part of Delivery 1 on Oct 5 2022. (DOI \url{https://archive.stsci.edu/doi/resolve/resolve.html?doi=10.17909/zqax-2y86})  This first delivery includes pre-JWST lens models, and fully-reduced, high-level HST, \textit{Spitzer}, and ALMA data. We briefly describe these data products in this section.

\subsubsection{\texorpdfstring{SGAS J1226$+$2152}{SGAS J1226+2152}}
We delivered reduced HST imaging data in ACS/F606W, F814W; and WFC3-IR/F110W, F140W and F160W. 
The details of the observation and data reduction are given in \citet{Sharon.2022}, and a map of the HST mosaic of the three main cluster cores that make up the foreground lens is given in Figure~1 of that paper. 
The Channel 1 and Channel 2 \textit{Spitzer} data reduction process is described in \citet{Florian.2021}. 

We provided two versions of strong lens model outputs for this cluster. Each model package contains the deflection, magnification, convergence, and shear maps, aligned to the same WCS solution as the HST imaging, for the best-fit model as well as a suite of 100 models taken from the MCMC sampling of the parameter space, for the purpose of estimating uncertainties. 
Version 1 (V1) is our ``best effort’’ pre-JWST lensing analysis, which is based all the existing HST imaging, and all the available spectroscopic information from the literature. The model is described in \citet{Sharon.2022}. 
For completeness, we also make public a previous version of the lens model, V0, which was used in several publications (Tejos et al. 2021, Dai et al.  2020, Solimano et al. 2021, Solimano et al. 2022), and is described in Tejos et al. (2021).

\subsubsection{\texorpdfstring{SGAS1723$+$34}{SGAS1723+34}}
We delivered reduced HST imaging data in WFC3-UVIS/F390W, F775W and WFC3-IR/F110W,F160W, and Channel 1 and Channel 2 \textit{Spitzer} data.  \citet{Sharon.2020} describe the HST data, data reduction, and strong lensing analysis.  The \textit{Spitzer} data are described in \citet{Florian.2021}, who also use this lens model in their analysis of the physical properties of the lensed galaxy.

\subsubsection{\texorpdfstring{SPT0418$-$47 and SPT2147$-$50}{SPT0418-47 and SPT2147-50}}
We provided reduced HST imaging data in WFC3-IR/F140W of both fields, described in \citet{Ma.2015}. ALMA data of both targets are described in \citet{Spilker.2016}. SPT0418$-$47 has continuum data at rest-frame 120 $\mu$m, 160 $\mu$m, and 380 $\mu$m, and SPT2147$-$50 has data at rest-frame 160 $\mu$m, 300 $\mu$m, 380 $\mu$m, and 450 $\mu$m. Both also have extensive sub/millimeter spectroscopy from the same datasets.

\section{Final thoughts}\label{sec:finalthoughts}
This paper describes TEMPLATES, a JWST Early Release Science program that was designed to optimize the study of galaxies with the JWST integral field units, by studying four very bright lensed galaxies.  We intend this paper to serve as the definitive description of the observations, the data reduction methods, the  Jupyter python notebooks that document our reduction steps, and the high-level data product deliverables.

We ask that papers using TEMPLATES data cite this overview paper.
Further, we ask that when data are reduced following the procedures described in this paper, that the resulting papers should cite this paper, as well as follow the citation guidelines in our Jupyter notebooks.

JWST is a transformative telescope that exceeds its (high) design expectations \citep{Rigby.2023}, and has powerful multiplexed spectroscopic capabilities.  It is therefore reasonable to expect that the most impactful discoveries from JWST, especially discoveries based on spectroscopic data, are still to come, as the scientific community learns how to fully exploit these complex datasets.  We look forward to discoveries about the nature of star formation in galaxies from the TEMPLATES dataset.  We share the data reduction methods and code we have developed for TEMPLATES (\S\ref{sec:dataredux} and \S\ref{sec:deliverables}), and lessons learned including recommendations for users (\S\ref{sec:lessonslearned}), in the hopes that our efforts are broadly helpful to other researchers as they transform JWST datasets into published results.

\begin{acknowledgments}
This work is based on observations made with the NASA/ESA/CSA JWST. The data were obtained from the Mikulski Archive for Space Telescopes at the Space Telescope Science Institute, which is operated by the Association of Universities for Research in Astronomy, Inc., under NASA contract NAS 5-03127 for JWST. These observations are associated with program \# 1355. 
The authors acknowledge that we developed this observing program with a zero-exclusive-access period.
This work was supported in part by a Student-Innovative-Creative-Hands-on Project (SICHOP) grant awarded by the Ohio Space Grant Consortium.
We especially thank James Muzerolle Page, Bernie Ruscher, and Ian Wong, for helping us understand the NIRSpec residual detector noise issue.  We thank our program coordinator Beth Perriello, and our science instrument reviewers Alberto Noriega-Crespo, Martha Boyer, and Alaina Henry.  We thank Richard Shaw for showing us how to download the data via script.
Support for JWST program 1355 was provided by NASA through a grant from the Space Telescope Science Institute, which is operated by the Association of Universities for Research in Astronomy, Inc., under NASA contract NAS 5-03127.
We are grateful for the collective contributions of the approximately 20,000 humans around the world who designed, built, tested, commissioned, and operate JWST.
\end{acknowledgments}

\bibliographystyle{aasjournal}
\bibliography{papers}

% Input tables
\begin{deluxetable}{llll}
% \tabletypesize{\scriptsize}
\tablecolumns{3}
\tablewidth{0pc}
\tablenum{1}
\label{tab:targets}
\tablecaption{Target list}
\tablehead{\colhead{short target name} & \colhead{full target name} & \colhead{RA (J2000)} & \colhead{DEC (J2000)}}
\startdata
SPT0418$-$47    & SPT-S J041839$-$4751.8   & 04 18 39.6790  & $-$47 51 52.68\\
SGAS1723$+$34   & SGAS~J1723$+$3411        & 17 23 36.4060  & $+$34 11 54.69\\
SGAS1226$+$21   & SGAS~J122651.3$+$215220  & 12 26 51.2960  & $+$21 52 19.97\\
SPT2147$-$50    & SPT$-$S J214720-5035.9   & 21 47 19.0120  & $-$50 35 54.50\\
\enddata
\tablecomments{Coordinates are for the center of the NIRSpec IFS pointing.
}
\end{deluxetable}

%
%  Updated version by TAH, that is not rotated and includes the references underneath each specific measurement value instead.  12/2023
%
%
\begin{deluxetable}{lcccc}
\tablecolumns{5}
\tablenum{2}
\label{tab:sampleprops}
\tablecaption{Properties of the targets}
\tablehead{
\colhead{property}  & \colhead{SGAS1723$+$34} & \colhead{SGAS1226$+$21} & \colhead{SPT2147$-$50} & \colhead{SPT0418$-$47}}
\startdata
~\vspace{-3mm}\\ % spacing
kind of galaxy  &   LBG    & LBG     & SMG    & SMG   \\
~\vspace{-1.5mm}\\ % spacing
%
% redshift
\multirow{2}{*}{redshift}    & $1.3293 \pm 0.0002$     & $2.9252 \pm 0.0009$     & $3.7604 \pm 0.0002$    & $4.2246 \pm 0.0004$    \\
& {\footnotesize \citet{Rigby.2021}} &    {\footnotesize \citet{Rigby.2018kvh}} &  {\footnotesize \citet{Reuter.2020}} &  {\footnotesize \citet{Reuter.2020}} \\
~\vspace{-1.5mm}\\ % spacing
%
% magnification
\multirow{2}{*}{magnification [$\mu$]}       & $52.7^{+3.3}_{-1.2}$     & $95 \pm 15$              & $6.6 \pm 0.4$          & $29.5 \pm 1.2$         \\
& {\footnotesize \citet{Florian.2021}} &  {\footnotesize \citet{Sharon.2022}} & {\footnotesize \citet{Spilker.2016}} &  {\footnotesize \citet{Cathey.2023}} \\
~\vspace{-1.5mm}\\ % spacing
%
% r_E
\multirow{2}{*}{$r_E$ (\arcsec)}     &  $4.8$                  & $6.5$                     &  $1.195 \pm 0.006$     & $1.207 \pm 0.002$     \\
& {\footnotesize This work} &  {\footnotesize This work} & {\footnotesize \citet{Spilker.2016}} & {\footnotesize \citet{Cathey.2023}}\\
~\vspace{-1.5mm}\\ % spacing
%
% Mstellar
\multirow{3}{*}{M$_*$ (\Msun)}  & ($5.95 ^{+2.2}_{-1.86}$) $\times10^{8}$ & $(1.46 \pm 0.34) \times10^{9}$     & ($6.1 \pm 1.9$) $\times 10^{10}$ & ($1.53 \pm 0.31$) $\times10^{10}$  \\
&  \multirow{2}{*}{\footnotesize \citet{Florian.2021}} & {\footnotesize \citet{Wuyts.2012} with} &  \multirow{2}{*}{\footnotesize This work} &   \multirow{2}{*}{\footnotesize \citet{Cathey.2023}}\\
& & {\footnotesize \citet{Sharon.2022} $\mu$} & \\
~\vspace{-1.5mm}\\ % spacing
%
% SFR
\multirow{3}{*}{SFR (\Msun~$yr^{-1}$)} & $8.19^{+0.41}_{-0.39}$         & $20.2 \pm 7.0$                      &  $520 \pm 80$       &  $128 \pm 19$       \\
& \multirow{2}{*}{\footnotesize \citet{Florian.2021}} &  {\footnotesize \citet{Wuyts.2012} with} & \multirow{2}{*}{\footnotesize This work} &  \multirow{2}{*}{\footnotesize \citet{Cathey.2023}} \\
& & {\footnotesize \citet{Sharon.2022} $\mu$} & \\
~\vspace{-1.5mm}\\ % spacing
%
% sSFR
\multirow{2}{*}{sSFR (Gyr$^{-1}$)}    & $13.8 \pm 5.1$                     &  $13.8 \pm 5.8$                     &   $8.5 \pm 3.0$                   & $8.4 \pm 2.1$                   \\
& {\footnotesize This work} &  {\footnotesize This work} & {\footnotesize This work} &  {\footnotesize \citet{Cathey.2023}} \\
~\vspace{-1.5mm}\\ % spacing
%
% A_V
\multirow{2}{*}{$A_V$}               & 0--0.5                   & 0.4                     &   $2.7 \pm 0.2$        & $3.8 \pm 0.1$        \\
& {\footnotesize \citet{Florian.2021}} &  {\footnotesize \citet{Chisholm.2019}} &  {\footnotesize This work} &  {\footnotesize \citet{Cathey.2023}} \\
~\vspace{-3mm}\\ % spacing
\enddata
\tablecomments{Properties of the TEMPLATES targets: type (Lyman break galaxy (LBG) or submillimeter galaxy (SMG), redshift $z$, lensing magnification $\mu$, Einstein radius $r_E$, stellar mass $M_*$, star formation rate SFR, specific star formation rate sSFR, and attenuation $A_V$. The Einstein radius for each of the SGAS targets is measured as the radius of a circle with area the same area as enclosed within the tangential critical curve for the source redshift. 
Underneath each measurement is the reference directly associated with it.
}
\end{deluxetable}

 % fancy latex version by TAH

\begin{deluxetable}{lllllll}
\tabletypesize{\scriptsize}
\rotate
\tablecolumns{7}
\tablewidth{0pc}
\tablenum{3}
\label{tab:obslog}
\tablecaption{JWST Observation Log for ERS Program TEMPLATES (PID 1355)}
\tablehead{\colhead{Observation} & \colhead{Target} & \colhead{Template} & \colhead{execution time (hours)} & \colhead{Start time UT} & \colhead{Settings} & \colhead{Note}}
\startdata
\cutinhead{Successful observations}
16 &	SPT0418$-$47  & NIRCam Imaging & 1.00 &  Aug 11, 2022 16:05:20 & F115W, F150W, F200W, F277W, F356W, F444W &  \\		
15 &	SPT0418$-$47  & MIRI Imaging   & 2.10 &  Aug 22, 2022 05:27:35 & F560W, F770W, F1000W, F1280W, F1500W, F1800W, F2100W &  \\		
11 &	SPT0418$-$47  & NIRSpec IFS    & 3.16 &  Oct 7, 2022 03:15:45  & G395M/F290LP   & \\	   
12 &	sky           & NIRSpec IFS    & 2.53 &  Oct 7, 2022 05:29:45  & G395M/F290LP   & B\\
13 &	SPT0418$-$47  & MIRI MRS       & 2.72 &  Jul 27, 2022 11:12:21 & setting B      &  A\\
17 &	SPT0418$-$47  & MIRI MRS       & 2.67 &  Aug 8, 2022 04:45:24  & setting C      & A\\
\hline
3 &	SGAS1226$+$21 & NIRCam Imaging & 0.94 &	Dec 16, 2022 11:28:55      & F115W, F150W, F200W, F277W, F356W, F444W & \\		
4 &	SGAS1226$+$21 &	MIRI Imaging   & 1.33 &  Jun 21, 2022 00:55:07     & F560W, F770W, F1000W, F1280W, F1500W, F1800W, F2100W &  \\		
1 &	SGAS1226$+$21 &	NIRSpec IFS    & 3.82 &  Dec 31, 2022 11:37:48     & G235H/F170LP   &  \\		
2 &	sky	      &  NIRSpec IFS   & 3.10 &  Dec 31, 2022 14:20:08         & G235H/F170LP   & B\\
5 &	SGAS1226$+$21 &	MIRI MRS       & 1.64 &  Dec 16, 2022 12:36:06     & setting C      &  \\	
6 &	sky	      &  MIRI MRS      & 1.50 &  Dec 16, 2022 13:52:56         & setting C, with on-source F560W simultaneous imaging      & B\\	
\hline                                         
9 &	SGAS1723$+$34 &	NIRCam Imaging & 1.38 &  Jun 28, 2022 10:18:15      & F115W, F150W, F200W, F277W, F356W, F444W\\		
10 &	SGAS1723$+$34 &	MIRI Imaging   & 1.79 &  Jun 29, 2022 23:37:44  & F560W, F770W, F1000W, F1280W, F1500W, F1800W, F2100W &  \\		
27 &	SGAS1723$+$34 &	NIRSpec IFS    & 5.67 &  Aug 31, 2022 04:57:03  & G140H/F100LP, G395H/F290LP   &  \\
28 &	sky	      &  NIRSpec IFS   & 5.04 &  Aug 31, 2022 09:25:42      & G140H/F100LP, G395H/F290LP   & B\\
\hline                                         
24 &	SPT2147$-$50  & NIRCam Imaging & 0.99 &  Sep 5, 2022 21:07:35   & F200W, F277W, F356W, F444W   &  \\ 
23 &	SPT2147$-$50  &	MIRI Imaging   & 2.03 &  Sep 7, 2022 01:55:02   & F560W, F770W, F1000W, F1280W, F1500W, F1800W, F2100W &  \\
19 &	SPT2147$-$50  &  NIRSpec IFS   & 2.78 &  Oct 16, 2022 04:02:49  & G395M/F290LP   & \\		
20 &	sky	          &  NIRSpec IFS   & 2.16 &  Oct 16, 2022 06:04:41  & G395M/F290LP   &  B \\
%
%
%\cutinhead{Pending observations}
21  &	SPT2147$-$50 & MIRI MRS	& 1.77  & Jun 9, 2023 10:12:56 & setting B &  \\
22  &	sky	         & MIRI MRS	& 1.10	& Jun 9, 2023 12:05:40 & setting B, with on-source F1000W simultaneous imaging & B \\
25  &	SPT2147$-$50 & MIRI MRS	& 1.25	& Jun 12, 2023 06:23:11 & setting C & \\
26  &	sky	         & MIRI MRS	& 1.25	& Jun 12, 2023 07:56:06 & setting C, with on-source F560W simultaneous imaging & B \\
\cutinhead{Failed observations, later re-observed successfully}
7  &	SGAS1723$+$34 &	NIRSpec IFS    & 9.88 &  Jul 15, 2022 13:39:18  & G140H/F100LP, G395H/F290LP  & A; C\\
\enddata
\tablecomments{JWST observations are ordered by status (succeeded or failed), then target, then observing mode.
  Note A: Observation includes dedicated offset background.
  Note B:  Observation is a dedicated offset background.
  Note C:  Rescheduled by WOPR 88493 as observation 27 in this program.
}
\end{deluxetable}

% also needed for moving affiliations to the end of doc
\allauthors

\end{document}